\documentclass[12pt]{article}

\usepackage{scicite}
\pdfoutput=1

\usepackage{graphics}
\usepackage{color}
\usepackage{graphicx}
\usepackage{amssymb}
\usepackage{wasysym}
\usepackage{bm}

\newenvironment{sciabstract}{%
\begin{quote} \bf}
{\end{quote}}

\newenvironment{methods}{%
    \section*{Methods}%
    \setlength{\parskip}{0pt}%
    }{}

\newcommand{\lesco}{La$_{1.7}$Eu$_{0.2}$Sr$_{0.1}$CuO$_{4}$}
\newcommand{\lnsco}{La$_{1.48}$Nd$_{0.4}$Sr$_{0.12}$CuO$_{4}$}

\newcounter{lastnote}

\renewcommand{\figurename}{{\bf{Fig.}}}

\makeatletter
\makeatletter \renewcommand{\fnum@figure}{{\bf{\figurename~\thefigure}}}
\makeatother

\title{Vortex phase diagram and the normal state of cuprates with charge and spin orders}

\author
{Zhenzhong Shi,$^{1\dag}$ P. G. Baity,$^{1,2\dag\dag}$ T. Sasagawa,$^{3}$ Dragana Popovi\'{c}$^{1,2\ast}$\\
\\
\normalsize{$^{1}$National High Magnetic Field Laboratory, Florida State University,}\\
\normalsize{Tallahassee, Florida 32310, USA}\\
\normalsize{$^{2}$Department of Physics, Florida State University,}\\
\normalsize{Tallahassee, Florida 32306, USA}\\
\normalsize{$^{3}$Materials and Structures Laboratory, Tokyo Institute of Technology,}\\
\normalsize{Kanagawa 226-8503, Japan}\\
\\
\normalsize{$^{\dag}$ Present address: Department of Physics, Duke University,}\\
\normalsize{Durham, North Carolina 27708, USA}\\
\normalsize{$^{\dag\dag}$ Present address: James Watt School of Engineering, University of Glasgow,}\\
\normalsize{Glasgow, G12 8QQ, Scotland, United Kingdom}\\
\normalsize{$^\ast$To whom correspondence should be addressed; E-mail: dragana@magnet.fsu.edu}
}

\date{}

\begin{document} 

\baselineskip=24pt

\maketitle 

\begin{sciabstract}
The phase diagram of underdoped cuprates in a magnetic field ($\bm{H}$) is the key ingredient in understanding the anomalous normal state of these high-temperature superconductors.  However, the upper critical field ($\bm{H_{c2}}$) or the extent of superconducting phase with vortices, a type of topological excitations, and the role of charge orders that are present at high $\bm{H}$, remain under debate.  We address these questions by studying stripe-ordered La-214, i.e. cuprates in which charge orders are most pronounced and zero-field transition temperatures $\bm{T_{c}^{0}}$ are lowest; the latter opens a much larger energy scale window to explore the vortex phases compared to previous studies.  By combining linear and nonlinear transport techniques sensitive to vortex matter, we determine the $\bm{T}$--$\bm{H}$ phase diagram, directly detect $\bm{H_{c2}}$, and reveal novel properties of the high-field ground state.  Our results  demonstrate that, while the vortex phase diagram of underdoped cuprates is not very sensitive to the details of the charge orders, quantum fluctuations and disorder play a key role as $\bm{T\rightarrow 0}$.  The presence of stripes, on the other hand, seems to alter the nature of the anomalous normal state, such that the high-field ground state is a metal, as opposed to an insulator.
\end{sciabstract}

\noindent\textbf{\large{Introduction}}\\
Cuprates are type-II superconductors\cite{Blatter1994,Giamarchi2009,LeDoussal2010,Rosenstein2010}: above the lower critical field $H_{c1}$, an external perpendicular magnetic field penetrates the material in the form of a solid lattice of vortices or quantized magnetic flux lines.  Since this vortex lattice is set in motion by the application of a current, the pinning of vortices by disorder ensures the zero-resistivity property of a superconductor below $T_{c}(H)$.  The disorder also turns the vortex lattice into a Bragg glass\cite{Giamarchi2009,LeDoussal2010}, i.e. a state that ``looks'' nearly as ordered as a perfect solid, but with many metastable states and the dynamics of a glass.  This phase melts into a vortex liquid (VL) when the temperature is high enough, or into an amorphous vortex glass (VG) for strong enough disorder.  The latter transition, in particular, can occur at low $T$ as a function of $H$, which increases the density of vortices and the relative importance of disorder\cite{Giamarchi2009,LeDoussal2010}.  In two-dimensional (2D) systems, the VG freezes at $T_g=0$, and thus the VG has zero resistivity only at $T_c=0$; signatures of a glassy, viscous VL, however, can be observed at low enough $T$, i.e. above $T_g$.  In general, the VL persists up to the crossover line $H_{c2}(T)$, i.e. up to the upper critical field $H_{c2}\equiv H_{c2}(T=0)$ where the superconducting (SC) gap closes.  Hence, the disorder suppresses the quantum melting field of the vortex lattice to be lower than $H_{c2}$.  Since the presence of disorder plays a crucial role in the vortex physics and leads to glassy dynamics, a method of choice for probing the energy landscape of such systems is the response to a small external force.  In particular, the key signatures of a Bragg glass and a glassy VL are nonlinear voltage-current ($V$--$I$) characteristics at low excitation currents: for $T<T_c(H)$, a vortex lattice or a Bragg glass has zero resistivity in the $I\rightarrow 0$ limit and a finite critical current; in contrast, the $I\rightarrow 0$ (i.e. linear) resistivity of a glassy VL is non-zero at all $T>0$, but its $V$--$I$ remains non-Ohmic.  At high enough $T$, the effects of disorder are no longer important and the $V$--$I$ of a (non-glassy) VL becomes Ohmic.  Although the vortex matter in cuprates has been studied extensively\cite{Blatter1994,Giamarchi2009,LeDoussal2010,Rosenstein2010}, the regime of high $H$ and low $T$, in which disorder and quantum effects are expected to dominate, has remained largely unexplored; in particular, there have been no $V$--$I$ measurements.  However, it is precisely this regime that is relevant for the determination of $H_{c2}$.  

In cuprates, which have a quasi-2D nature, the value of $H_{c2}$ continues to be of great interest, because the strength of pairing correlations is an essential ingredient in understanding what controls the value of $T_{c}^{0}$, the nature of the pairing mechanism, and the pseudogap regime\cite{Keimer2015}.  However, the interpretation of experimental data has been controversial and contradictory [\cite{Yu2016,Griss2014,Cyr2017,Julien2017} and refs. therein].  At least a part of the controversy lies in a disagreement on whether the VL regime vanishes\cite{Griss2014} or persists\cite{Yu2016} at $T=0$, although it has been shown that the VL may survive up to surprisingly high fields\cite{Shi2014,Yu2016} at $T\ll T_{c}^{0}$.  For example, in highly underdoped La$_{2-x}$Sr$_{x}$CuO$_{4}$ (LSCO), in which charge order is at best very weak\cite{Jacobsen2015}, the vortex phase diagram was proposed based on the evidence\cite{Shi2014} for quantum criticality of the linear resistivity associated with $T=0$ transitions from the vortex lattice to VG, and then from the VG to insulator, i.e. a 2D superconductor-insulator transition (SIT)\cite{Fisher1990}, at higher $H$.  Charge orders, however, are ubiquitous in all hole-doped cuprates, in a range of dopings centered near $x=1/8$, and their interplay with superconductivity at high $H$ is one of the key open questions\cite{Julien2015,Comin2016}.  To explore the possible effects of charge order on the phases of vortex matter and the nature of the normal state, we study the in-plane resistivity ($\rho_{ab}$) of other underdoped cuprates from the La-214 family because, at $H=0$, they exhibit short-range charge orders with strongest correlations.  Furthermore, they do not have the complication present in YBa$_2$Cu$_3$O$_{6+x}$ (YBCO), in which an additional, longer-range charge order emerges at high $H$ \cite{Julien2015,Comin2016}.  We also employ nonlinear ($V$--$I$) transport as a direct probe of vortex matter.  The measurements extend over an unusually large, more than three-orders-of-magnitude-wide range of $T$ down to $T/T_{c}^{0} \lesssim 0.003$ (Methods).

We focus on La$_{1.8-x}$Eu$_{0.2}$Sr$_x$CuO$_4$ (LESCO) and La$_{1.6-x}$Nd$_{0.4}$Sr$_x$CuO$_4$ (LNSCO), in which charge order coexists with the antiferromagnetic spin-density-wave order\cite{Fradkin2015} at low enough $T<T_{SO}<T_{CO}$; here $T_{SO}$ and $T_{CO}$ are the onsets of static, short-range spin and charge orders or ``stripes'', respectively, and $T_{c}^{0}(x)$ is suppressed for doping near $x=1/8$ (Supplementary Fig.~1).  As shown below, the low values of $T_{c}^{0}$ in this regime allow us to reveal and probe deep into the normal state by applying $H$ up to 35~T.  Fields perpendicular to CuO$_2$ planes, i.e. along the $c$ axis, enhance both spin\cite{Lake2002,Chang2008} and charge orders\cite{Wen2012,Huecker2013}.  These effects are observed only below $T_{c}^{0}$, not above, and in samples away from $x=1/8$, such as $x\sim 0.10$, where the stripe order is weaker.  The emergence and an enhancement of the charge order by $H$ below $T_{c}^{0}$, accompanying the suppression of high-temperature superconductivity, has been observed\cite{Gerber2015} also in YBCO.  While there is no evidence of coincident static spin order in YBCO \cite{Keimer2015}, it has been argued\cite{Badoux2016} that the charge order in YBCO is similar to that in La-based cuprates and, in fact, that charge-density-wave modulations persist in the $T=0$ field-induced normal state in underdoped YBCO and in Hg- and La-based cuprates\cite{Badoux2016}.  Thus, the stripe order in LESCO and LNSCO is expected to persist even after $T_c(H)$ is suppressed to zero.

We find that the vortex phase diagrams of La-214 underdoped cuprates are qualitatively the same regardless of the presence or strength of the charge orders: as $T\rightarrow 0$, the vortex lattice is separated from the high-field ground state by a wide range of fields where quantum phase fluctuations and disorder (i.e. a viscous VL) dominate.  On general grounds, the same conclusions should apply also to other cuprates below optimum doping.  By establishing the $T$--$H$ phase diagram over an unprecedented range of parameters and demonstrating that measurements, such as nonlinear transport, are needed down to $T\ll T_{c}^{0}$ to determine $H_{c2}$, our work also resolves the lasting controversy in the cuprate research.  Furthermore, we reveal novel properties of the high-field normal (non-SC) state in the presence of stripes: the insulatinglike, $\ln(1/T)$ dependence of the in-plane resistivity, the origin of which has been a long-standing puzzle, is suppressed by $H$, suggesting a metallic high-field ground state.  This is in contrast to the early work\cite{Ando1995} in the absence of static stripes, where $H$ does not seem to affect the $\ln(1/T)$ behavior, and the high-field ground state is an insulator.\\
\vspace*{-12pt}

\noindent\textbf{\large{Results}}\\
\noindent\textbf{In-plane resistivity of La$\bm{_{1.8-x}}$Eu$\bm{_{0.2}}$Sr$\bm{_{x}}$CuO$\bm{_{4}}$ and La$\bm{_{1.6-x}}$Nd$\bm{_{0.4}}$Sr$\bm{_{x}}$CuO$\bm{_{4}}$.}
Our samples were single crystals with the nominal composition \lesco\, and\linebreak \lnsco\, (Methods).  $T_{c}^{0}$ was defined as the temperature at which the linear resistance $R_{ab}\equiv\lim_{I_{dc}\rightarrow 0} V/I$ (that is, $\rho_{ab}$) becomes zero.  For \lesco, $T_{c}^{0}=(5.7\pm 0.3)$~K, $T_{SO}\sim 15$~K, $T_{CO}\sim 40$~K (Supplementary Fig.~1A), and the pseudogap temperature $T_{pseudogap}\sim 175$~K \cite{Cyr2017}.  For \lnsco, $T_{c}^{0}=(3.6\pm 0.4)$~K, $T_{SO}\sim 50$~K, $T_{CO}\sim 70$~K (Supplementary Fig.~1B), and $T_{pseudogap}\sim 150$~K \cite{Cyr2017}. 

Figure~\ref{MR}A shows the in-plane magnetoresistance (MR) measurements at fixed $T$ for \lesco\, (see Supplementary Fig.~2A for similar data on \lnsco).  The positive MR, indicative of the suppression of superconductivity, appears below $\sim 35$~K.  However, as $T$ decreases further, a negative MR develops at the highest fields, resulting in a peak in $\rho_{ab}(H)$ at $H=H_{peak}(T)$ for $T< T_{c}^{0}$.  In cuprates, only positive MR has been discussed so far, with the exception of a broad MR maximum\cite{Ando1995} in underdoped LSCO.   
Its similarities to that in 2D films of conventional superconductors near a field-tuned SIT have led to the suggestion\cite{Steiner2005} that, in both cases, $H_{peak}$ represents the field scale above which the SC gap, a measure of the pairing strength between electrons that form Cooper pairs, vanishes.  

Figure \ref{MR}B (Supplementary Fig.~2B for LNSCO) shows the $\rho_{ab}(T)$ curves extracted from the MR measurements for $H\leq 18$~T; the MR data at higher $H$ are discussed further below.  When the normal state sheet resistance, $R_{\square/{\mathrm{layer}}}$, is close to the quantum resistance for Cooper pairs, $R_{Q}=h/(2e)^{2}$, which occurs for $T\approx 15$~K, the $\rho_{ab}(T)$ curves start to separate from each other: lower-$H$ $\rho_{ab}(T)$ curves exhibit a  metalliclike drop associated with superconductivity, while higher-$H$ curves show a tendency towards insulating behavior.  This is also remarkably reminiscent of the behavior of 2D films of many materials at a $T=0$  SIT \cite{Goldman2010,Gantmakher2010}, where the critical resistance is close to $R_Q$.  Such a transition is generally attributed to quantum fluctuations of the SC phase and, hence, loss of phase coherence; Cooper pairs that form the SC condensate thus survive the transition to the insulator.  Although underdoped cuprates are highly anisotropic, layered materials, and thus behave effectively as 2D systems\cite{Emery1995,Shi2014,Baity2016}, these similarities between \textit{bulk} systems, LESCO and LNSCO, and 2D films are striking: it took only specially engineered, double-layer transistors to observe\cite{Bollinger2011} the critical resistance of $R_Q$ for the $H=0$, electric-field-driven SIT in LSCO.  These similarities thus suggest the importance of phase fluctuations in LESCO and LNSCO at low $T$.    
Figure~\ref{MR}B shows that the increase of $H$ clearly leads to the suppression of $T_c(H)$ and, just like in highly underdoped LSCO \cite{Shi2014}, to a peak in $\rho_{ab}(T)$ that shifts to lower $T=T_{peak}(H)$.  In LSCO, $T_{peak}(H)$ was attributed\cite{Shi2014} to the onset of the viscous VL regime.  The values of $T_{c}(H)$ (see also Supplementary Fig.~3), $H_{peak}(T)$, and $T_{peak}(H)$ are shown in Fig.~\ref{phase} over a wide range of $T$ and $H$ for both materials.

The quantum melting of the vortex lattice, in which $\rho_{ab}=0$ as the vortices are pinned by disorder\cite{Blatter1994,Giamarchi2009,LeDoussal2010,Rosenstein2010},  occurs when $T_c(H)\rightarrow 0$, e.g. for $\sim 5.5$~T in LESCO (Fig.~\ref{phase}A).  On general grounds, in type-II superconductors the vortex lattice melts into a vortex liquid or glass, i.e. a regime of strong phase fluctuations.  Indeed, at the lowest $T$ in the intermediate $H$ regime, the data are described best with the power-law fits ${\rho}_{ab}(H,T) = {\rho}_{0}(H)T^{{\alpha}(H)}$ in both materials (Supplementary Fig.~4), suggesting a true SC state ($\rho_{ab}=0$) only at $T=0$ when the vortices are frozen, and consistent with the expectations for a viscous VL above its glass freezing temperature\cite{Dorsey1992,VGTg-1,VGTg-2} $T_g=0$.  This finding is similar to that\cite{Shi2014} in highly underdoped LSCO ($x=0.06, 0.07$).  Here we employ also nonlinear transport measurements to probe the vortex matter directly.\\
\vspace*{-12pt}

\noindent\textbf{Nonlinear transport and superconducting correlations.}
The second technique, therefore, involves measurements of $V$--$I$ characteristics at fixed $H$ and $T$ (Methods), in addition to the linear resistance $R_{ab}$ discussed above.  When $T<T_c$, $dV/dI$ is zero as expected in a superconductor (Fig.~\ref{IV}A), since small values of $I_{dc}$ are not able to cause the depinning of the vortex lattice.  However, at higher $H$, where $T_c$ is suppressed to zero, a zero-resistance state is not observed even at $I_{dc}=0$ (Fig.~\ref{IV}A), down to the lowest $T$ (Fig.~\ref{IV}B), but the $V$--$I$ characteristic remains non-Ohmic and $dV/dI$ increases with $I_{dc}$.  This type of behavior is indeed expected from the motion of vortices in the presence of disorder, i.e. it is a signature of a glassy, viscous VL\cite{Giamarchi2009,LeDoussal2010,Rosenstein2010}.  At higher $H$ and $T$, the non-Ohmic response vanishes (Fig.~\ref{IV}).  

A comprehensive study of nonlinear transport (see, e.g., Supplementary Fig.~5) over the entire range of $T$ and $H$ was performed on \lesco.  The non-Ohmic behavior was established for all $H<H^{\ast}(T)$ and Ohmic behavior for $H>H^{\ast}(T)$ in Fig.~\ref{phase}A; as $T\rightarrow 0$, $H^{\ast}$  extrapolates to $\sim 20$~T.   We note the quantitative agreement between $H^{\ast}(T)$, the boundary of the viscous VL obtained from nonlinear transport, with the values of $T_{peak}(H)$ obtained from the linear resistivity measurements.  Thus $T_{peak}(H)$ can be indeed used to identify the extent of the viscous VL.  Moreover, at $T<0.4$~K, both $H^{\ast}(T)$ and $T_{peak}(H)$ agree, within the error, with $H_{peak}(T)$, suggesting that, for $T\ll T_{c}^{0}$, $H_{peak}$ may be indeed interpreted as the field scale corresponding to the closing of the SC gap\cite{Steiner2005}.  In \lnsco, $T_{peak}(H)$ and $H_{peak}(T)$ are also found to be the same, within the error, at low enough $T<0.7$~K (Fig.~\ref{phase}B); $V$--$I$ measurements at low $T$ (Supplementary Fig.~6), confirm that transport remains non-Ohmic, for example, up to $H^{\ast}=(22\pm 3)$~T at $T\sim 0.05$~K, comparable to $H_{peak}(T\sim 0.05$~K$)\sim 27$~T, and well beyond the quantum melting field of the vortex lattice ($\sim 4$~T). 

The extent of SC fluctuations observed in linear transport was determined from the positive MR at high $T$ (Fig.~\ref{MR}A) as the field $H_{c}'(T)$ above which the MR increases as $H^2$ (Supplementary Fig.~7), as expected in the high-$T$ normal state.  As in other studies on cuprates\cite{Rullier-Albenque2007,Rullier-Albenque2011,Rourke2011,Shi2013,Shi2014,Baity2016}, the result can be fitted with $H_{c}' = H_{0}'[1-(T/T_{2})^{2}]$ (Fig.~\ref{phase}A).  Remarkably, as $T\rightarrow 0$, both $H_{c}'$ and $H^{\ast}$ extrapolate to $\sim 20$~T in \lesco, consistent with $H^{\ast}(T=0)$ being the depairing field, i.e. $H_{c2}$.  Region III in Fig.~\ref{phase} ($H>H^{\ast}, H_{c}'$) then corresponds to the $H$-induced normal state.  Similar, albeit fewer measurements of $H_{c}'$ in \lnsco\, (Fig.~\ref{phase}B) are consistent with this picture.\\
\vspace*{-12pt}

\noindent\textbf{High-field normal state.}
The highest field $\rho_{ab}(T)$ data are shown in Fig.~\ref{normal}.  In \lesco, $\rho_{ab}\propto\ln(1/T)$ is observed in this regime over a temperature range of one-and-a-half decades, without any sign of saturation down to at least $\sim 0.07$~K or $T/T_{c}^{0}\sim 10^{-2}$ (Fig.~\ref{normal}A).  The $\ln(1/T)$ behavior was first reported\cite{Ando1995} in underdoped La$_{2-x}$Sr$_x$CuO$_4$;  it was not dependent on $H$, and its origin is still not well understood\cite{Shi2014}.  Nevertheless, experiments suggest that the  high-$H$ normal-state in underdoped La$_{2-x}$Sr$_x$CuO$_4$ is an insulator\cite{Ando1995,Shi2014,Karpinska1996}.  In contrast, a clear weakening of the insulatinglike, $\ln(1/T)$ behavior with $H$ is observed in \lesco\, (Fig.~\ref{normal}A), strongly suggesting that $\rho_{ab}$ becomes independent of $T$, i.e. metallic, at much higher fields $\gtrsim 55$~T (Fig.~\ref{normal}A inset).   

Similar $\rho_{ab}\propto\ln(1/T)$ behavior is found in \lnsco\, (Fig.~\ref{normal}B), in agreement with earlier experiments\cite{Adachi2005} performed up to 15~T and down to 1.5~K.  However, while that work reported\cite{Adachi2005} that $\rho_{ab}$ became independent of the field strength for $H\geqslant 11$~T, lower $T$ and higher $H$ have allowed us to reveal the weakening of the $\ln(1/T)$ behavior by $H$ also in \lnsco\, (Fig.~\ref{normal}B).  The apparent saturation at low $T$ is attributed to the presence of $T_{peak}(H)$, which moves to lower $T$ with increasing $H$.  The onset of metallic behavior is anticipated at $\sim 70$~T (Fig.~\ref{normal}B inset).  The results thus strongly suggest that the high-field ground state of striped cuprates is a metal.

We note that the possibility that SC fluctuations persist in the $H>H^{\ast}$ regime (e.g. beyond $H^{\ast}\sim 20$~T as $T\rightarrow 0$ in \lesco) cannot be completely ruled out based on these measurements, as $H_{c}'(T)$ may acquire a ``tail'' at low $T$ such that the fitted $H_{0}'<H_{c}'(T=0)$ \cite{Shi2014}.   However, magnetotransport studies that employ also parallel magnetic fields do confirm the absence of any observable remnants of superconductivity for $H>H^{\ast}(T\rightarrow 0)$; those results will be presented elsewhere.  In any case, it would be interesting to perform additional studies of this peculiar high-field normal state that is characterized by the $\ln(1/T)$ temperature dependence and the negative MR, as well as to extend the measurements to $H>55$~T to probe the properties of the anticipated high-field metal phase. \\
\vspace*{-12pt}

\noindent\textbf{Vortex phase diagram.}
Figure~\ref{phase} shows that, in $H=0$, phase fluctuations become dominant at $T\lesssim 15$~K~$\sim (2$-$3)T_{c}^{0}$ in \lesco,  and at $T\lesssim 20$~K~$\sim 6\,T_{c}^{0}$ in \lnsco, i.e. about an order of magnitude below $T_{pseudogap}$.  Gaussian fluctuations of amplitude and phase become observable below $\sim 35$~K in \lesco, i.e. $\sim (33$-$48)$~K in \lnsco\, as in previous work\cite{Adachi2005}.  Therefore, we find no evidence for pairing at temperatures comparable to $T_{pseudogap}$, similar to the conclusions of other, recent studies\cite{Baity2016,Cyr2017}.  Measurements at relatively high $T<T_c$ suggest low values of $H^{\ast}$ needed to close the gap (Fig.~\ref{phase}), consistent with some reports\cite{Chang2012a,Cyr2017} of low values of $H_{c2}$.  For example, it was reported\cite{Chang2012a} that $H_{c2}\sim8$~T on \lesco\, with a similar $T_{c}^{0}$, but those experiments were performed at fairly high $T$ (see Supplementary Fig.~8).  Figure~\ref{phase}A shows that $H^{\ast}\sim 8$~T at a comparably high $T\sim 1$~K.  By extrapolating to $T\rightarrow 0$ in different cuprate families, it was then argued\cite{Griss2014} that there is no VL at $T=0$.  However, as our study demonstrates, it is only by systematically tracking the evolution of SC correlations with $H$ and $T$ down to $T/T_{c}^{0}\ll 1$, that reliable $T\rightarrow 0$ extrapolations can be made.  Indeed, Fig.~\ref{phase} shows clearly that the regime of viscous VL \textit{broadens} dramatically as $T$ is reduced: the VL thus survives up to fields $H_{c2}=H^{\ast}\sim 20$~T in \lesco\, ($\sim 25$~T in \lnsco) as $T\rightarrow 0$, i.e. well beyond the quantum melting field of the vortex lattice ($H_m\sim 5.5$~T and $H_m\sim 4$~T, respectively, for \lesco\, and \lnsco).  The relatively high values of $H_{c2}$ are consistent with the $H=0$ opening of the SC gap at $T$ several times higher than $T_{c}^{0}$ (albeit $\ll T_{pseudogap}$).

The existence of a broad phase fluctuations regime in $H=0$ can be attributed to several factors.  The first one is the effective 2D nature of the materials, which leads to the Berezinskii-Kosterlitz-Thouless transition and broadening of $\rho_{ab}(T)$ close to $T_{c}^{0}$ \cite{Baity2016}.  The second one is the presence of stripes, i.e. the \textit{intrinsic} nanoscale phase separation or intrinsic electronic inhomogeneity, which can result from competing interactions even in the absence of disorder\cite{Schmalian2000,Sami2015}: the presence of stripes in cuprates has been suggested to lead to an intrinsically granular SC state\cite{Kapitulnik2019}.  Indeed, in contrast to highly underdoped LSCO \cite{Shi2014}, the $\rho_{ab}(T)$ curves of striped LESCO and LNSCO exhibit a two-step development of superconductivity (Supplementary Figs.~2-4, 9B) similar to novel hybrid structures with engineered nanoscale phase separation\cite{Kessler2010,Eley2012,Han2014,Chen2018}, granular films of conventional superconductors\cite{Parendo2007,Gantmakher2010,CIQPT}, relatively clean heavy-fermion\cite{Park2012} and iron-based superconductors\cite{Xiao2012} in the presence of competing orders, and some novel 2D materials\cite{Sajadi2018}.  Finally, even \textit{homogeneous} disorder may lead to the inhomogeneity of the pairing amplitude\cite{Bouadim2011} on the scale of the coherence length, forming SC islands weakly coupled by the Josephson effect; although inhomogeneous, the pairing amplitude is finite throughout the system, and the SC transition occurs at $T_{c}^{0}$ because of the loss of phase coherence.  Spatially inhomogeneous pairing gaps persisting above $T_{c}^{0}$ have been observed in cuprates, such as Bi$_2$Sr$_2$CaCu$_2$O$_{8+\delta}$ [e.g. \cite{Pasupathy2008}], and in films of conventional superconductors\cite{Kamplapure2013}, including on both sides of the SIT\cite{Sherman2012}.

Interestingly, in films with weak and homogeneous disorder, the application of $H$ itself can lead to the emergent granularity of the SC state\cite{Ganguly2017}.  Generally speaking, since increasing $H$ increases the effective disorder\cite{Giamarchi2009,LeDoussal2010}, as $T\rightarrow 0$ it also leads to the suppression of $T_c(H)$, i.e. the suppression of the ordered, vortex lattice phase (region I in Fig.~\ref{phase}) to fields below $H_{c2}$.  This results in an intermediate regime characterized by large quantum phase fluctuations, i.e. a viscous VL in Fig.~\ref{phase}, that extends up to $H_{c2}$.  Therefore, consistent with general expectations\cite{Blatter1994,Giamarchi2009,LeDoussal2010,Rosenstein2010,CIQPT,TVojta2013}, $H^{\ast}(T\rightarrow 0)=H_{c2}$ in Fig.~\ref{phase} corresponds to the upper critical field of the \textit{clean} system, i.e. in the absence of disorder.  

Our results (Fig.~\ref{phase}) indicate that, in analogy to 2D SC films near the $H$-tuned SIT\cite{Goldman2010,Gantmakher2010,CIQPT}, the MR peak in LESCO and LNSCO allows access to both bosonic (positive MR) and fermionic (negative MR) regimes.  While $H_{peak}(T)$ tracks $H^{\ast}(T)$ and $T_{peak}(H)$ at low temperatures, there is a clear bifurcation at higher $T$.  This results in the $H^{\ast}(T)<H<H_{peak}(T)$ bosonic regime in which, in contrast to the viscous VL, the transport is Ohmic.  Therefore, we identify this regime as the (Ohmic) vortex liquid.  This is consistent with the finding, based on other types of probes, that a line similar to $T_{peak}(H)$ separates the VL from the viscous VL in the high-temperature ($T<T_{c}^{0}$), low-$H$ regime of several cuprate families\cite{LeDoussal2010,Rosenstein2010}.  It is still an open question whether that line represents a true second-order glass transition or just a crossover energy scale.  The nonmonotonic behavior of $H_{peak}(T)$ observed at low $T$, however, is a new and unexpected finding.  It suggests that, for certain $H$, the system can pass from a regime dominated by SC phase fluctuations (i.e. a VL) to a normal, albeit very unusual, state, and then again to a phase-fluctuations-dominated regime (i.e. a viscous VL) as $T$ is lowered.  The understanding of this intriguing behavior, however, is beyond the scope of this study.
\\
\vspace*{-12pt}

\noindent\textbf{\large{Discussion}}\\
Our key results are summarized in the $T$-$H$ phase diagrams (Fig.~\ref{phase}) of striped LESCO and LNSCO, which have been mapped out over more than three orders of magnitude of $T$ and deep into the $H$-induced normal state using two different, complementary techniques that are sensitive to global phase coherence.  In particular, by relying only on transport techniques, we have been able to obtain a self-consistent set of data points and achieve quantitative agreement of our results, allowing us to compare different energy scales.  Furthermore, by using cuprates with a very low $T_{c}^{0}$, we have been able to access a much larger energy scale window to explore the vortex phases compared to previous studies.  Indeed, Supplementary Fig.~8 shows that our measurements extend to effectively much lower $T$ and higher $H$ compared to previous studies of $H_{c2}$ on these and other underdoped cuprates.  This has allowed us to probe the previously inaccessible regimes, including deep into the normal state.  We have revealed that the unusual, insulatinglike behavior [$\rho_{ab}\propto\ln(1/T)$] of the normal state is suppressed by even higher $H$, leading to a metallic high-field ground state in striped cuprates.

Our findings have important implications for other cuprates, especially YBCO, which is considered the cleanest cuprate, but just like any other real material, it does contain some disorder\cite{Julien2017,Wu2016}.  Even if the disorder is weak, its effects will be amplified at low $T$ with increasing $H$, leading to the quantum melting of the vortex lattice into a VL below $H_{c2}$.  Therefore, probes sensitive to vortex matter, employed over a wide range of $T$ and $H$, and studying the response of the system to a \textit{small} external force, are needed to determine $H_{c2}$.  Indeed, for transport measurements, for example, Fig.~\ref{IV} demonstrates that, due to the intrinsically nonlinear nature of the $V$-$I$ characteristics of the vortex matter, using high excitations $I_{dc}$ would yield much higher measured values of $\rho_{ab}$ and even change the sign of its temperature dependence (Fig.~\ref{IV}B).  As a result, the observation of $T_{peak}(H)$ and the identification of the viscous VL regime, would not be possible.  In YBCO, low-$T$ magnetization measurements have reported\cite{Yu2016} the melting of the vortex lattice into a ``second vortex solid'' with a much weaker shear modulus, somewhat reminiscent of the viscous VL regime in Fig.~\ref{phase}, but $H_{c2}$ was not reached with the accessible fields.  Indeed, conflicting MR results, i.e. both a positive MR only\cite{Boeuf2011} and a peak\cite{Griss2016}, have been reported for the same material and doping, leaving the question of $H_{c2}$ in YBCO open.  Nonlinear transport studies at low excitations of YBCO with a very low $T_{c}^{0}$, comparable to that in our striped LESCO and LNSCO crystals or in highly underdoped LSCO \cite{Shi2014}, are thus needed to identify the vortex phases, determine $H_{c2}$, and probe the normal state.  

A comparison of spectroscopic data on a variety of hole-doped cuprates, in which the La-214 family was not considered, has established\cite{Hufner2008} that the SC gap $\sim 5k_{B}T_{c}^{0}$ with the value of $\sim$40~meV at optimal doping.  This indeed suggests inaccessibly high $H_{c2}$ ($\sim 100$~T), although this same relationship yields an order of magnitude lower gap $\sim$2--3~meV for $T_{c}^{0} \simeq 4$--6~K that are comparable to those in our LESCO and LNSCO crystals.  Similarly, in the La-214 family at optimal doping, where $T_{c}^{0}\sim 40$~K, it was found\cite{Lake2001,Wagman2016,Li2018} that the SC energy scale $\sim20$~meV.  It is thus reasonable to expect a SC gap $\sim2$~meV when $T_{c}^{0} \simeq 4$--6~K, i.e. of the same order as $H_{c2}$ found in our study.  

In \lesco\, and \lnsco, which have different strengths of stripe correlations, we have established qualitatively the same vortex phase diagrams as in highly underdoped LSCO\cite{Shi2014}, in which there is no clear evidence of charge order\cite{Jacobsen2015}.   The LSCO study\cite{Shi2014} was performed using a different method, as described above, and on samples grown by two different techniques.  Therefore, the qualitative agreement between the vortex phase diagrams obtained on three different materials confirms that our findings are robust, and that the presence of different phases of vortex matter in underdoped La-214 cuprates is not very sensitive to the details of the competing charge orders.  Our data also highlight the key role of disorder in understanding the $T\rightarrow 0$ behavior of underdoped cuprates, and demonstrate that the SC phase with vortices persists up to much higher fields $H_{c2}$ than those argued previously.  Even in conventional superconductors, the interplay of vortex matter physics, disorder, and quantum fluctuations leads to the enhancement of $H_{c2}$ as $T\rightarrow 0$, a long-standing puzzle in the field\cite{Sacepe2018}.  It should thus come as no suprise that the precise values of $H_{c2}$ may be also affected by the presence of stripes.  In particular, stronger stripe correlations in \lnsco\, than in \lesco\, seem to enhance the VL regime as $T\rightarrow 0$, but this issue is beyond the scope of this work.  On the other hand, since strong stripe correlations have not altered the vortex phase diagrams, there is no reason to expect that much weaker, at least in $H=0$, charge orders in other cuprates will modify the vortex phase diagrams qualitatively.  Our results thus strongly suggest that our conclusions should apply to all underdoped cuprates, as supported by the agreement with the spectroscopic data on other cuprates, including the La-214 family. 
However, whether the vortex liquid regime extends out to overdoped regions of the cuprate phase diagram where the (normal state) pseudogap closes remains an open question for future study.

\begin{methods}

\noindent\textbf{Samples.}  Several single crystal samples of La$_{1.8-x}$Eu$_{0.2}$Sr$_x$CuO$_4$ (Supplementary Fig.~1A) with a nominal $x=0.10$ and La$_{1.6-x}$Nd$_{0.4}$Sr$_x$CuO$_4$ (Supplementary Fig.~1B) with a nominal $x=0.12$ were grown by the traveling-solvent floating-zone technique\cite{Takeshita2004}.  From the X-ray fluorescence analysis using an X-ray analytical and imaging microscope (HORIBA XGT-5100), it was confirmed that the chemical compositions were close to the nominal values within the experimental error and spatially homogeneous with the standard deviation less than 0.003 for a $\sim 1$~mm$^2$ area mapping with $\sim\diameter 10~\mu$m resolution.  This is supported by the fact that the structural phase transition from the low-temperature orthorhombic to low-temperature tetragonal phase, which reflects the global chemical composition, was very sharp as observed in the temperature dependence of the $c$-axis resistivity, measured\cite{Baity2018} on a bar-shaped \lnsco\, sample with dimensions $0.24\times 0.41\times 1.46$~mm$^3$ ($a\times b\times c$) and $T_{c}^{0}=(3.51\pm 0.06)$~K.  SQUID and torque magnetometry measurements were performed on a bar-shaped \lesco\, crystal with dimensions $1.19\times 0.24\times 0.78$~mm$^3$ ($a\times b\times c$).  Detailed transport measurements were performed on two crystals, shaped as rectangular bars suitable for direct measurements of the in-plane resistance, with dimensions $3.06\times 0.37\times 0.53$~mm$^3$ and $3.82\times 1.19\times 0.49$~mm$^3$ ($a\times b\times c$) for \lesco\, and \lnsco, respectively.  Gold contacts were evaporated on polished crystal surfaces, and annealed in air at 700 $^{\circ}$C. The current contacts were made by covering the whole area of the two opposing sides with gold to ensure uniform current flow, and the voltage contacts were made narrow  to minimize the uncertainty in the absolute values of the resistance.  Multiple voltage contacts on opposite sides of the crystals were prepared.  The distance between the voltage contacts for which the data are shown is 1.53~mm for LESCO and 2.00~mm for LNSCO.  Gold leads ($\approx 25~\mu$m thick) were attached to the samples using the Dupont 6838 silver paste, followed by the heat treatment at 450\,$^{\circ}$C in the flow of oxygen for 15 minutes. The resulting contact resistances were less than 0.1~$\Omega$ for LESCO (0.5~$\Omega$ for LNSCO) at room temperature. \\
\vspace*{-12pt}

\noindent The values of $T_{c}^{0}$ and the behavior of the samples did not depend on which voltage contacts were used in the measurements, reflecting the absence of extrinsic (e.g. compositional) inhomogeneity in these crystals.  Furthermore, both $T_{c}^{0}$ and the sample behavior remained astonishingly stable with time (see, e.g., Supplementary Fig.~3), without which it would have not been possible to conduct such an extensive and systematic study that requires matching data obtained using different cryostats and magnets (see below) over the period of 2-3 years during which this study was performed, thus further attesting to the high quality of the crystals.  Even the value of $T_{c}^{0}=(3.6\pm 0.4)$~K in \lnsco\, indicates high sample homogeneity: first, it agrees with the value  ($\sim 4$~K) obtained\cite{Tranquada1997} from magnetic susceptibility, a bulk probe, and second, it is lower than the values of $T_{c}^{0}$ away from this doping at which the superconductivity is most suppressed [\cite{Ichikawa2000} and refs. therein; also Supplementary Fig.~1B].  In contrast, some of the other transport studies of \lnsco\, have found $T_{c}^{0}\sim 6$~K \cite{Adachi2005} and $T_{c}^{0}\sim 7$~K \cite{Chang2012a}, pointing to either a different effective doping or inhomogeneity in their samples.  Likewise, $T_{c}^{0}\approx 5.7$~K of our \lesco\, crystal agrees with the bulk susceptibility measurements (Supplementary Fig.~9A), while its high-$T$ resistivity (Supplementary Fig.~9B) is about five times, that is, significantly lower than in some other studies\cite{Chang2012a} of the same material.  Similarly, the high-$T$ resistivity of our \lnsco\, crystal is at least $\sim 30$\% lower than previously published values\cite{Adachi2005}, indicating that our samples are at least as homogeneous. The agreement between the resistive $T_c(H)$ and the irreversibility field $H_{irr}(T)$ obtained from torque magnetometry in LESCO (Supplementary Fig.~10) further confirms the absence of extrinsic inhomogeneity and high crystal quality.  In fact, it is probably because of the relatively low level of disorder that our striped LESCO and LNSCO crystals exhibit a two-step $\rho_{ab}(T)$ (Supplementary Figs.~2-4), known to reflect the onset of local and global superconductivity in various 2D systems\cite{Parendo2007,Gantmakher2010,CIQPT,Kessler2010,Eley2012,Han2014,Chen2018}, as discussed above.  A similar two-step $\rho_{ab}(T)$ is observed\cite{Adachi2005,Li2007,Ding2008,Chang2012a,Stegen2013} also in other La-214 cuprates near $x=1/8$.
\\
\vspace*{-12pt}

\noindent\textbf{Measurements.}   The standard four-probe ac method ($\sim 13$~Hz) was used for measurements of the sample resistance, with the excitation current (density) of 10~$\mu$A ($\sim 5 \times 10^{-3}$~A\,cm$^{-2}$ and $\sim 2 \times 10^{-3}$~A\,cm$^{-2}$ for LESCO and LNSCO, respectively).  $dV/dI$ measurements were performed by applying a dc current bias (density) down to 2~$\mu$A ($\sim 1 \times 10^{-3}$~A\,cm$^{-2}$ and $\sim 4 \times 10^{-4}$~A\,cm$^{-2}$ for LESCO and LNSCO, respectively) and a small ac current excitation $I_{ac}\approx 1~\mu$A ($\sim$ 13 Hz) through the sample, while measuring the ac voltage across the sample.  For each value of $I_{dc}$, the ac voltage was monitored for 300~s and the average value recorded.  The data that were affected by Joule heating at large dc bias were not considered.  In the high-field normal state, for example, the dc current bias where Joule heating becomes relevant, identified as the current above which the $V$-$I$ characteristic changes from Ohmic to non-Ohmic, was $I_{dc}>100~\mu$A at the lowest $T$; at higher $T$, such as 1.7~K, that current was $I_{dc}\gtrsim10$~mA.  In all measurements, a $\pi$ filter was connected at the room temperature end of the cryostat to provide a 5~dB (60~dB) noise reduction at 10~MHz (1~GHz). \\
\vspace*{-12pt}

\noindent The experiments were conducted in several different magnets at the National High Magnetic Field Laboratory: a dilution refrigerator (0.016 K $\leqslant$ T $\leqslant$ 0.7~K), a $^{3}$He system (0.3~K $\leqslant$ T $\leqslant$ 35~K), and a variable-temperature insert (1.7~K $\leqslant$ T $\leqslant$ 200~K) in superconducting magnets ($H$ up to 18~T), using 0.1 -- 0.2~T/min sweep rates; a portable dilution refrigerator (0.02 K $\leqslant$ T $\leqslant$ 0.7 K) in a 35~T resistive magnet, using 1~T/min sweep rate; and a $^{3}$He system (0.3 K $\leqslant$ T $\leqslant$ 20 K) in a 31~T resistive magnet, using 1 -- 2~T/min sweep rates.  Below $\sim0.06$~K, it was not possible to achieve sufficient cooling of the electronic degrees of freedom to the bath temperature, a common difficulty with electrical measurements in the mK range.  This results in a slight weakening of the $\rho_{ab}(T)$ curves below $\sim0.06$~K for \textit{all} fields.  We note that this does not make any qualitative difference to the phase diagram (Fig.~\ref{phase}).  The fields, applied perpendicular to the CuO$_{2}$ planes, were swept at constant temperatures. The sweep rates were low enough to avoid eddy current heating of the samples.  \\
\vspace*{-12pt}

\noindent The resistance per square per CuO$_2$ layer $R_{\square/\mathrm{layer}}=\rho_{ab}/l$, where $l=6.6$~\AA\, is the thickness of each layer.\\
\vspace*{-12pt}

\end{methods}

\bibliography{scibib}

\bibliographystyle{Science}

\noindent{\Large{\textbf{Acknowledgements}}}

\noindent We acknowledge helpful discussions with L. Benfatto, V. Dobrosavljevi\'c, and N. Trivedi.  \textbf{Funding:} This work was supported by NSF Grants Nos. DMR-1307075 and DMR-1707785, and the National High Magnetic Field Laboratory (NHMFL) through the NSF Cooperative Agreement No. DMR-1157490 and the State of Florida.  \textbf{Author contributions:} Single crystals were grown and prepared by T.S.; Z.S. and P.G.B. performed the measurements and analysed the data; D.P. contributed to the data analysis and interpretation; Z.S., P.G.B. and D.P. wrote the manuscript; D.P. planned and supervised the investigation.  All authors commented on the manuscript.  \textbf{Competing interests:} The authors declare that they have no competing interests.  \textbf{Data and materials availability:}  All data needed to evaluate the conclusions in the paper are present in the paper and/or the Supplementary Materials.  Additional data related to this paper may be requested from the authors.

\clearpage

%
\begin{figure}
\centerline{\includegraphics[width=0.70\textwidth]{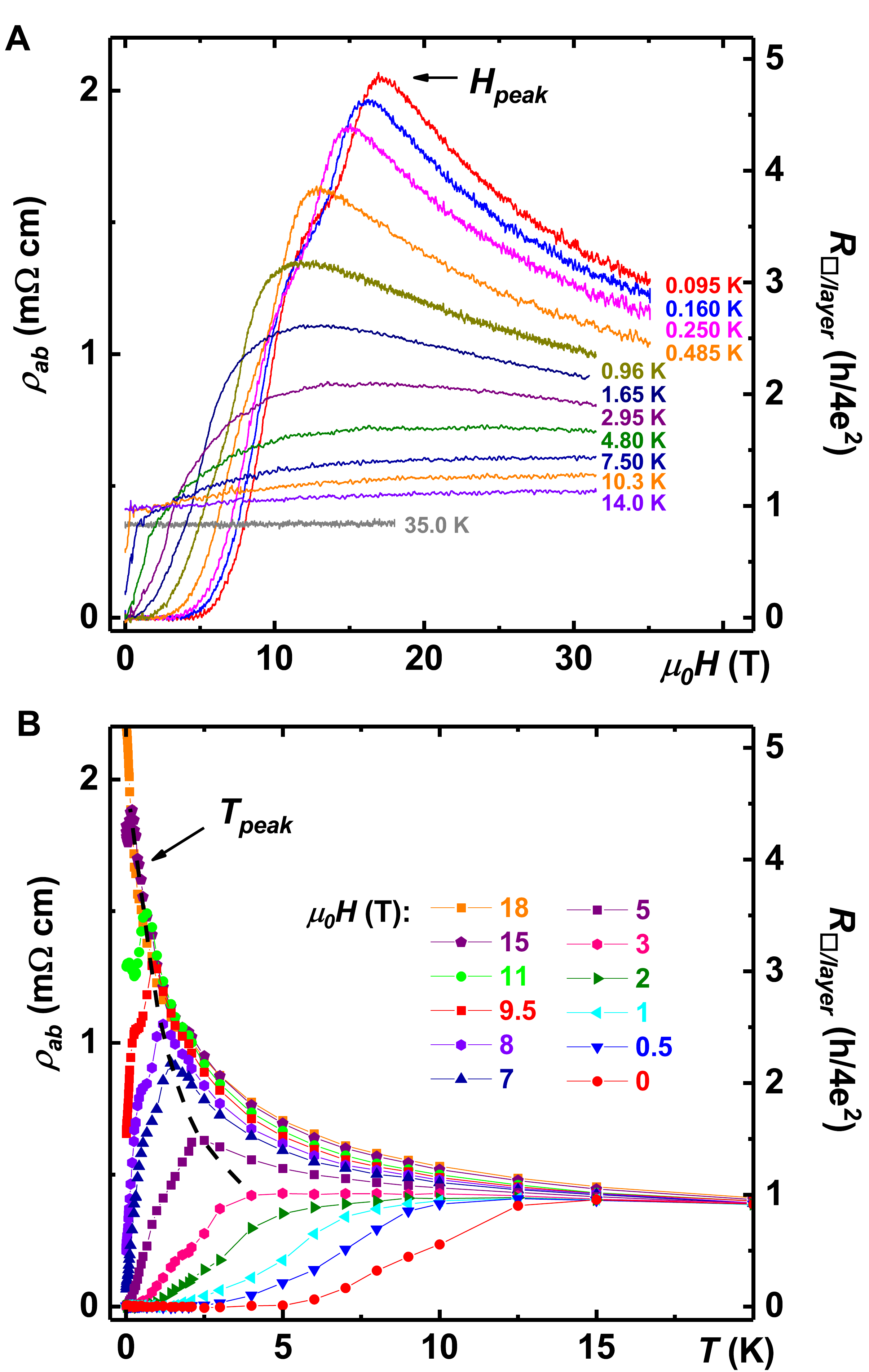}}
\caption{\textbf{The dependence of the in-plane resistivity $\bm{\rho_{ab}}$ of \lesco\, on magnetic field $\bm{H\parallel c}$ and $\bm{T}$.}  \textbf{A}, ${\rho}_{ab}(H)$ for several $T$ from 0.095 K to 35.0 K, as shown.  At low $T$, $\rho_{ab}(H)$ exhibits a strong peak at $H=H_{peak}(T)$.  The right axis shows the corresponding  $R_{\square/\mathrm{layer}}$ in units of quantum resistance for Cooper pairs, $R_{Q}=h/(2e)^{2}$.  \textbf{B}, $\rho_{ab}(T)$ for several $H\leq 18$~T, as shown.  Solid lines guide the eye.  The black dashed line tracks the values of $T_{peak}(H)$.  
}
\label{MR}
\end{figure}
%

%
\begin{figure}
\centerline{\includegraphics[width=1.05\textwidth]{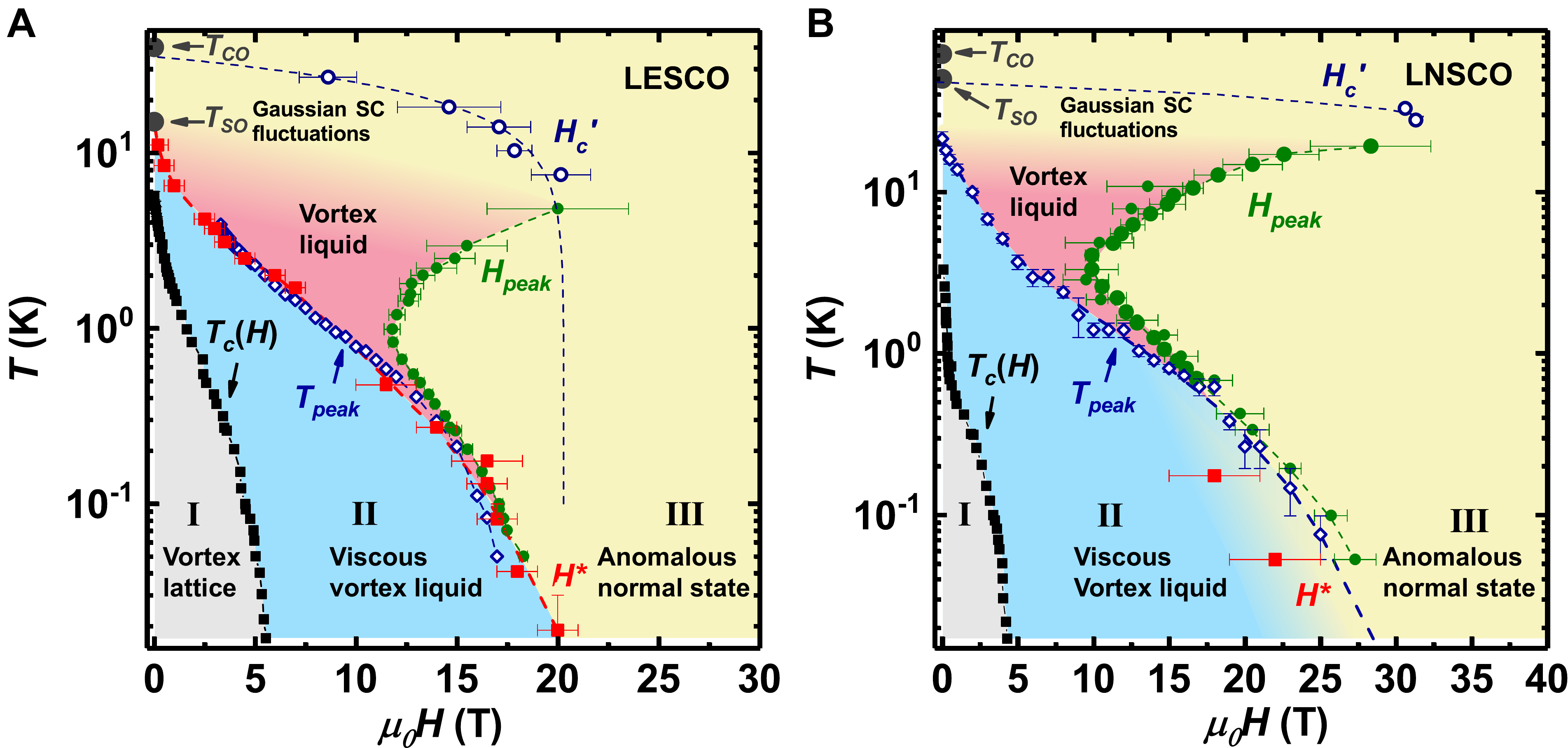}}
\caption{\textbf{In-plane transport $\bm{T}$--$\bm{H}$ phase diagram of striped cuprates with $\bm{H\parallel c}$ axis.}  
\textbf{A}, \lesco; \textbf{B}, \lnsco.  $T_c(H)$ (black squares) mark the boundary of the pinned vortex lattice, which is a superconductor with $\rho_{ab}=0$ for all $T<T_c(H)$ [region I; $T_c(H)>0$].  $H^{\ast}(T)$ symbols mark the boundary of the viscous vortex liquid, in which $dV/dI$ is non-Ohmic [for $H<H^{\ast}(T)$] and which freezes into a vortex glass at $T=T_c=0$; dashed red line guides the eye.  Ohmic behavior is found at $H>H^{\ast}(T)$.  $H^{\ast}(T=0)$ thus corresponds to the upper critical field $H_{c2}$.  $H_{peak}(T)$ (green dots) represent fields above which the MR changes from positive to negative.  The region $H^{\ast}<H<H_{peak}$, in which the MR is positive but transport is Ohmic, is identified as the vortex liquid.  $T_{peak}(H)$ (open blue diamonds) track the positions of the peak in $\rho_{ab}(T)$.  $H_{c}'(T)$ is the field above which SC fluctuations are not observed; Gaussian fluctuations of the SC amplitude and phase are expected at the highest $T$ and $H<H_{c}'(T)$.  The highest field region (III) corresponds to the $H$-induced normal state.  The dashed line in (\textbf{A}) is a fit with $\mu_{0}H_{c}'[$T$] = 20.3[1-(T[$K$]/35.4)^{2}]$, and the error bars indicate the uncertainty in $H_{c}'$ that corresponds to 1 SD in the slopes of the linear fits in Supplementary Fig.~7.  In (\textbf{B}), SC fluctuations vanish between 33~K and 48~K for $H=0$, and the dashed line is a guide to the eye.  Zero-field values of $T_{SO}$ and $T_{CO}$ are also shown; both spin and charge stripes are known to be enhanced by $H$ (see main text).  Except for $T_{c}(H)$, lines do not represent phase boundaries, but finite-temperature crossovers.
}
\label{phase}
\end{figure}
%

%
\begin{figure}
\centerline{\includegraphics[width=0.6\textwidth]{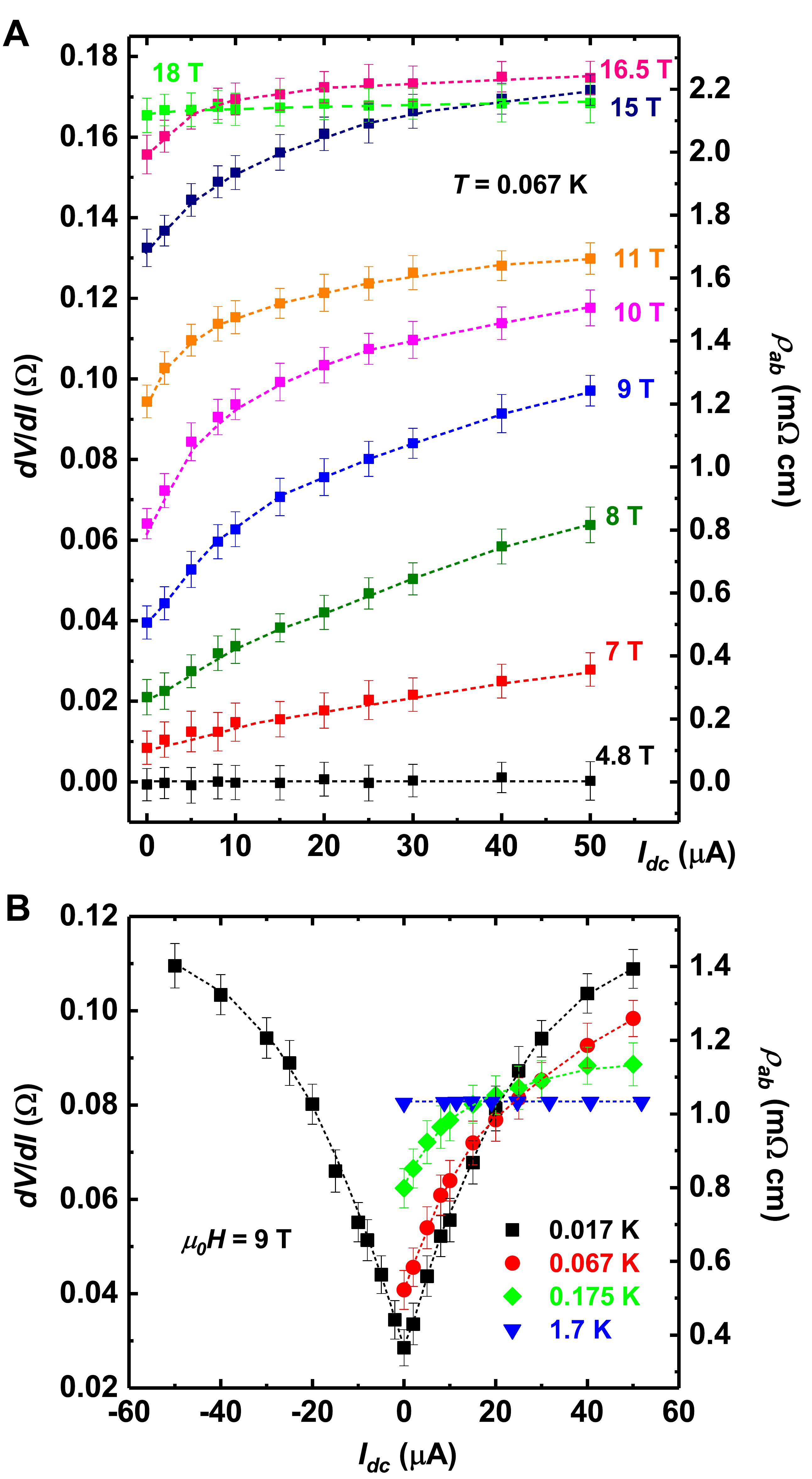}}
\caption{\textbf{Nonlinear in-plane transport in La$\bm{_{1.7}}$Eu$\bm{_{0.2}}$Sr$\bm{_{0.1}}$CuO$\bm{_4}$.}  \textbf{A,} Differential resistance $dV/dI$ as a function of dc current $I_{dc}$ for several $H\leq 18$~T at $T=0.067$~K.  In the bottom trace, for which $T<T_c(H=4.8$~T)$\approx 0.08$~K, $dV/dI$ is zero as expected in a superconductor.  \textbf{B}, $dV/dI$ vs. $I_{dc}$ for several $T$ at $H=9$~T.  The linear resistance ($dV/dI$ for $I_{dc}\rightarrow 0$) has a metalliclike temperature dependence but, at higher $I_{dc}>20~\mu$A, the temperature dependence of $dV/dI$ is insulatinglike.
Dashed lines guide the eye.
}
\label{IV}
\end{figure}
%

%
\begin{figure}
\centerline{\includegraphics[width=0.76\textwidth]{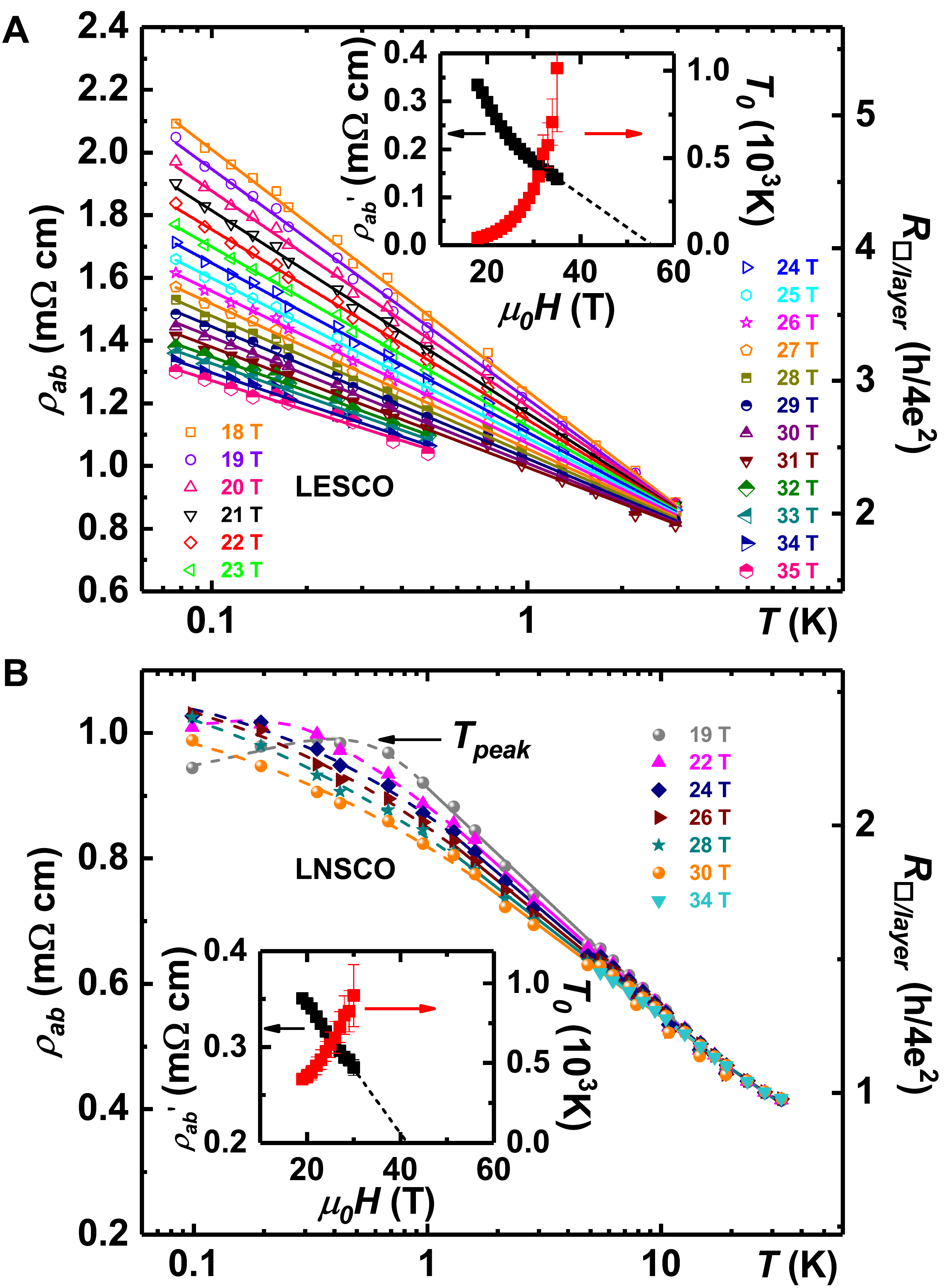}}
\caption{\textbf{Temperature dependence of the in-plane resistivity $\bm{\rho_{ab}}$ at the highest $\bm{H}$, i.e. in the normal state.}  
\textbf{A}, \lesco\, and \textbf{B}, \lnsco.  Solid lines are fits to $\rho_{ab}=\rho_{ab}'(H)\ln(T_0(H)/T)$, and dashed lines guide the eye.  The arrow in (\textbf{B}) shows $T_{peak}(H=19$~T$)$.  Insets: Fitting parameters $\rho_{ab}(H)$ and $T_0(H)$.  The decrease of the slopes $\rho_{ab}'$ (black squares) with $H$ indicates the weakening of the insulatinglike, $\ln(1/T)$ dependence with $H$. The linear extrapolation of $\rho_{ab}'$ to zero (dashed line) provides a rough estimate of the field $\sim 55$~T in \lesco, i.e. $\sim 71$~T in \lnsco, where insulatinglike behavior vanishes.  
}
\label{normal}
\end{figure}
%


\clearpage

\noindent\textbf{\Large{Supplementary Materials}}
\vspace*{12pt}

\makeatletter
\makeatletter \renewcommand{\fnum@figure}{{\bf{\figurename~S\thefigure}}}
\makeatother

\setcounter{figure}{0}

\baselineskip=24pt

%
\begin{figure}[h]
\centerline{\includegraphics[width=0.53\textwidth]{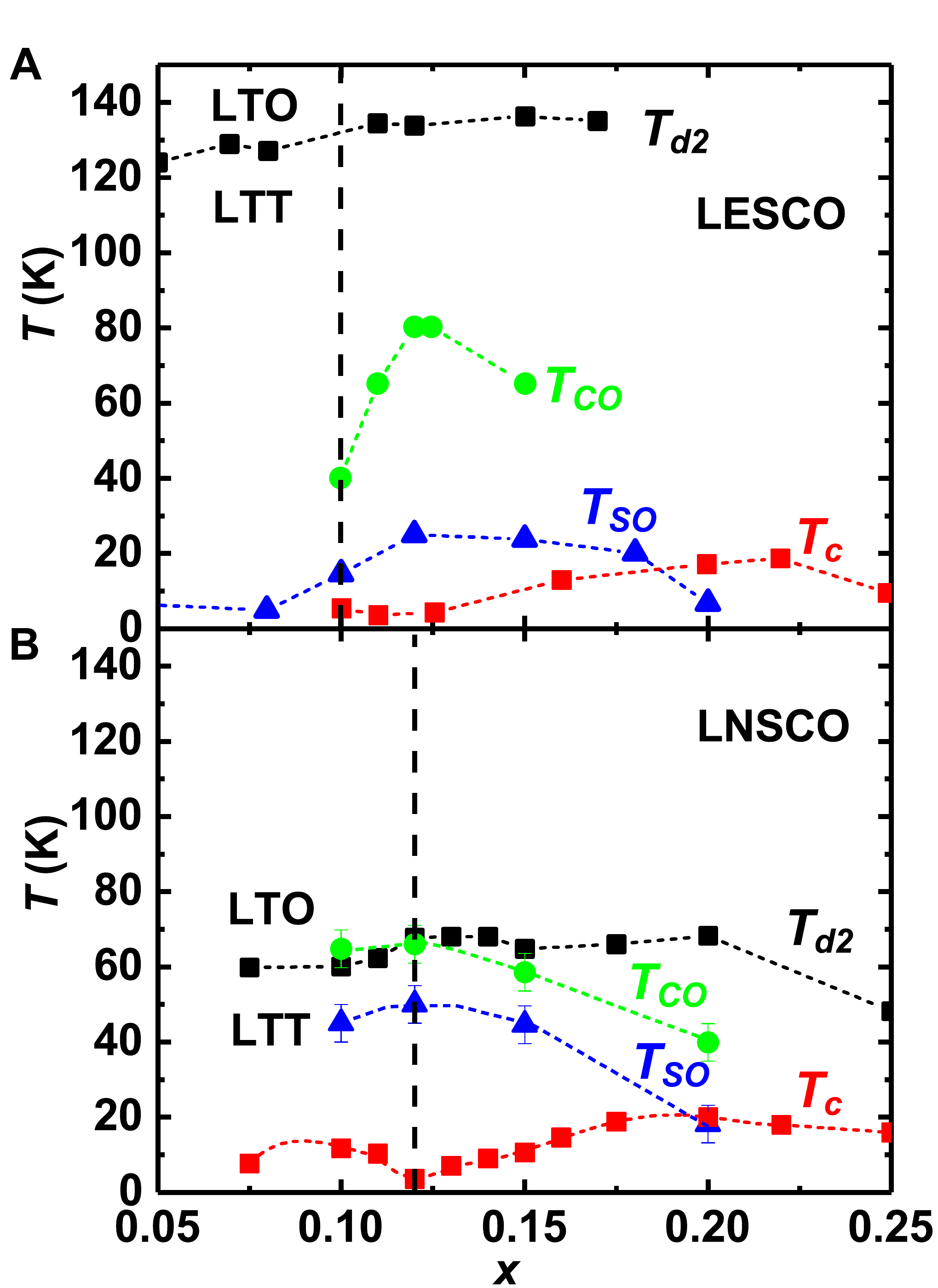}}
\caption{\textbf{$\bm{T}$--$\bm{x}$ phase diagram of striped cuprates for $\bm{H=0}$.}  \textbf{A}, La$_{1.8-x}$Eu$_{0.2}$Sr$_x$CuO$_4$ (LESCO); data are reproduced from refs.~[S1-S3]. \textbf{B}, La$_{1.6-x}$Nd$_{0.4}$Sr$_x$CuO$_4$ (LNSCO); data are reproduced from refs.~[S1, S4-S7].  
$ T_{d2} $ marks the transition temperature from the low-temperature orthorhombic (LTO) structure to a low-temperature tetragonal (LTT) structure.  $T_{CO} $ and $T_{SO} $ are the onset temperatures for charge and spin orders, respectively. $T_{c}$ is the superconducting transition temperature.  Dashed lines guide the eye.  The vertical dashed lines show the doping of our single crystals: $x=0.10$ for LESCO and $x=0.12$ for LNSCO.}
\label{FigS1}
\end{figure}

\clearpage

%
\begin{figure}
\centerline{\includegraphics[width=0.78\textwidth]{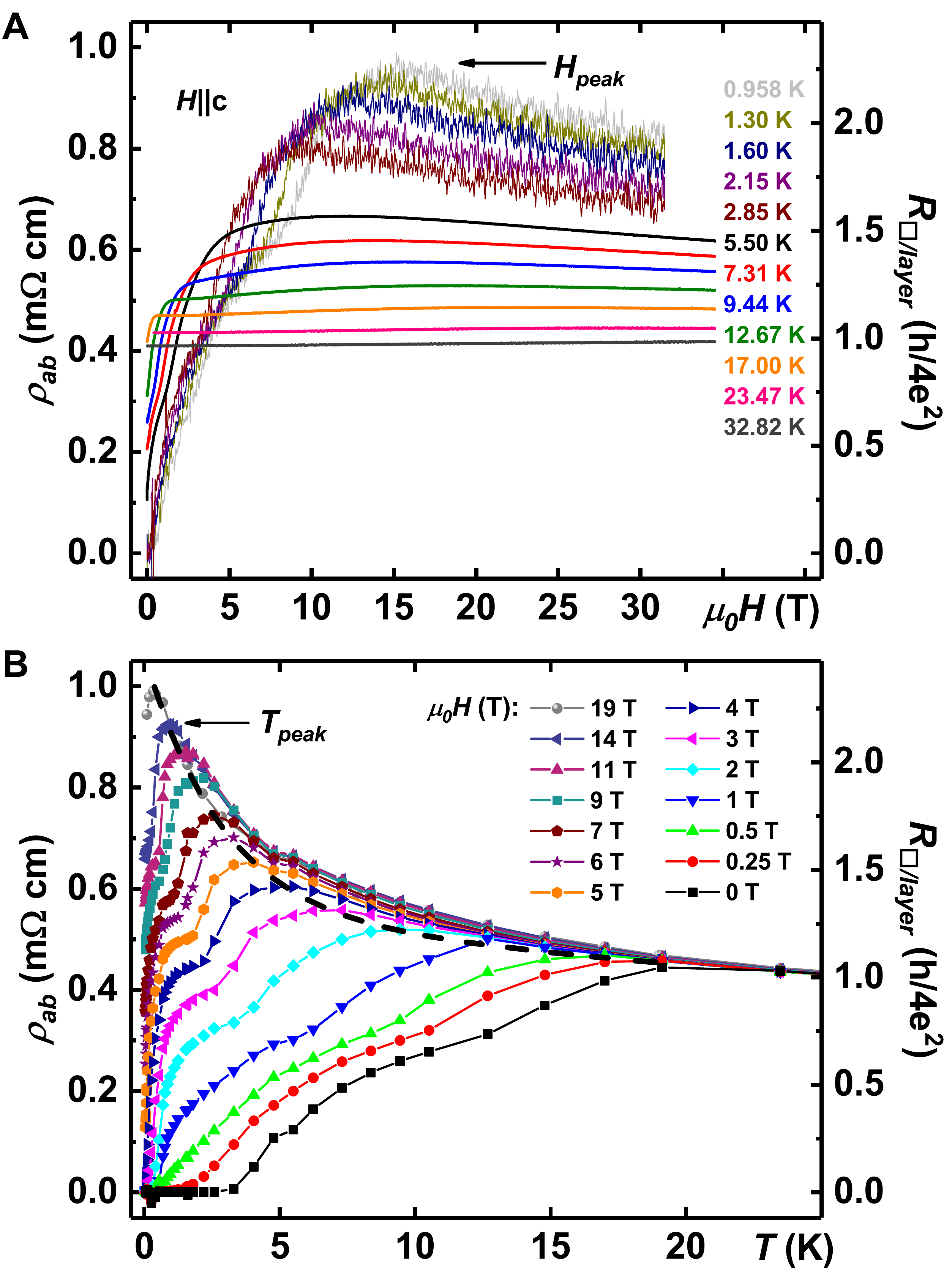}}
\caption{\textbf{The dependence of the in-plane resistivity $\bm{\rho_{ab}}$ of La$\bm{_{1.48}}$Nd$\bm{_{0.4}}$Sr$\bm{_{0.12}}$CuO$\bm{_4}$ on magnetic field $\bm{H\parallel c}$ and $\bm{T}$.}  \textbf{A}, ${\rho}_{ab}(H)$ for several $T$ from 0.095 K to 35.0 K, as shown.  At low $T$, $\rho_{ab}(H)$ exhibits a peak at $H=H_{peak}(T)$.  The right axis shows the corresponding  $R_{\square/\mathrm{layer}}$ in units of quantum resistance for Cooper pairs, $R_{Q}=h/(2e)^{2}$.  \textbf{B}, $\rho_{ab}(T)$ for several $H\leq 18$~T, as shown.  Solid lines guide the eye.  The black dashed line tracks the values of $T_{peak}(H)$.  The splitting of the $\rho_{ab}(T)$ curves for different $H$ becomes pronounced when $R_{\square/\mathrm{layer}}\approx R_{Q}$, at $T\sim 20$~K.
}
\label{MR-LNSCO}
\end{figure}
%

\clearpage

\begin{figure}
\centerline{\includegraphics[width=\textwidth]{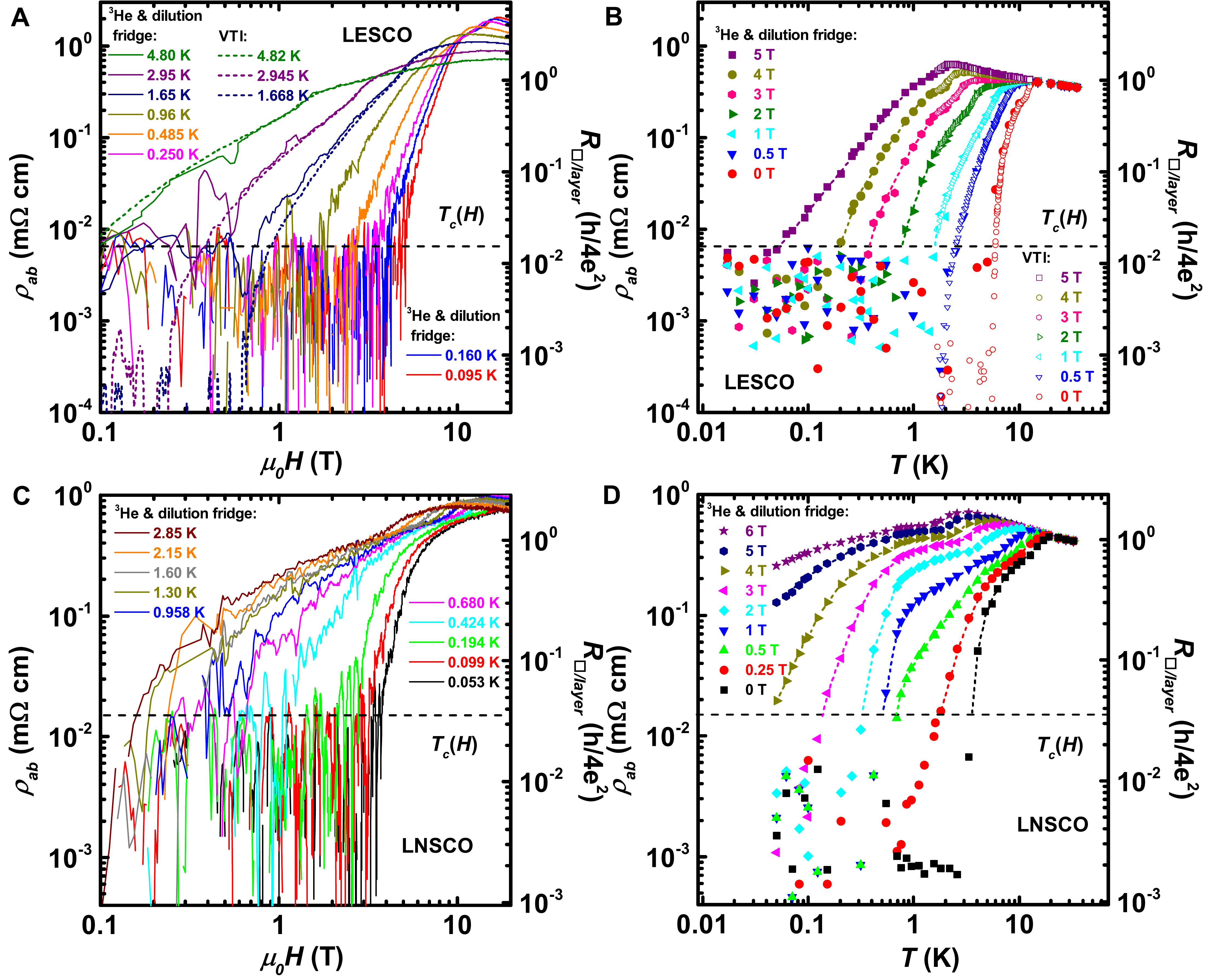}}
\caption{\textbf{Determination of the zero-resistance $\bm{T_c(H)}$.} 
In a magnified, i.e. log-log plot of $\rho_{ab}(H)$ curves obtained in resistive magnets at fixed $T$ (\textbf{A} and \textbf{C}, for \lesco\, and \lnsco, respectively), the field where the resistivity is smaller than the experimental noise floor is defined as the ``zero-resistance field'' at a given $T$.  That temperature, in turn, is identified as the zero-resistance $T_c(H)$ in this field.  The values of $T_c(H)$ are also shown in the log-log plots of $\rho_{ab}(T)$ at fixed $H$, measured in superconducting magnets (\textbf{B} and \textbf{D}, for \lesco\, and \lnsco, respectively).  The noise floor was 0.5~m$\Omega$ in resistance, corresponding to $\rho_{ab}$ of $6.5\times 10^{-3}$~m$\Omega$cm for LESCO and $1.5\times 10^{-2}$~m$\Omega$cm for LNSCO, as indicated by horizontal dashed lines.  As shown, measurements were also performed in several different cryostats: a $^{3}$He system, a dilution refrigerator, and a variable-temperature insert (VTI).  The agreement between data obtained during different runs illustrates the remarkable stability of the crystals with time, further attesting to their high quality.
}
\label{Tc}
\end{figure}

\clearpage

\begin{figure}
\centerline{\includegraphics[width=0.8\textwidth]{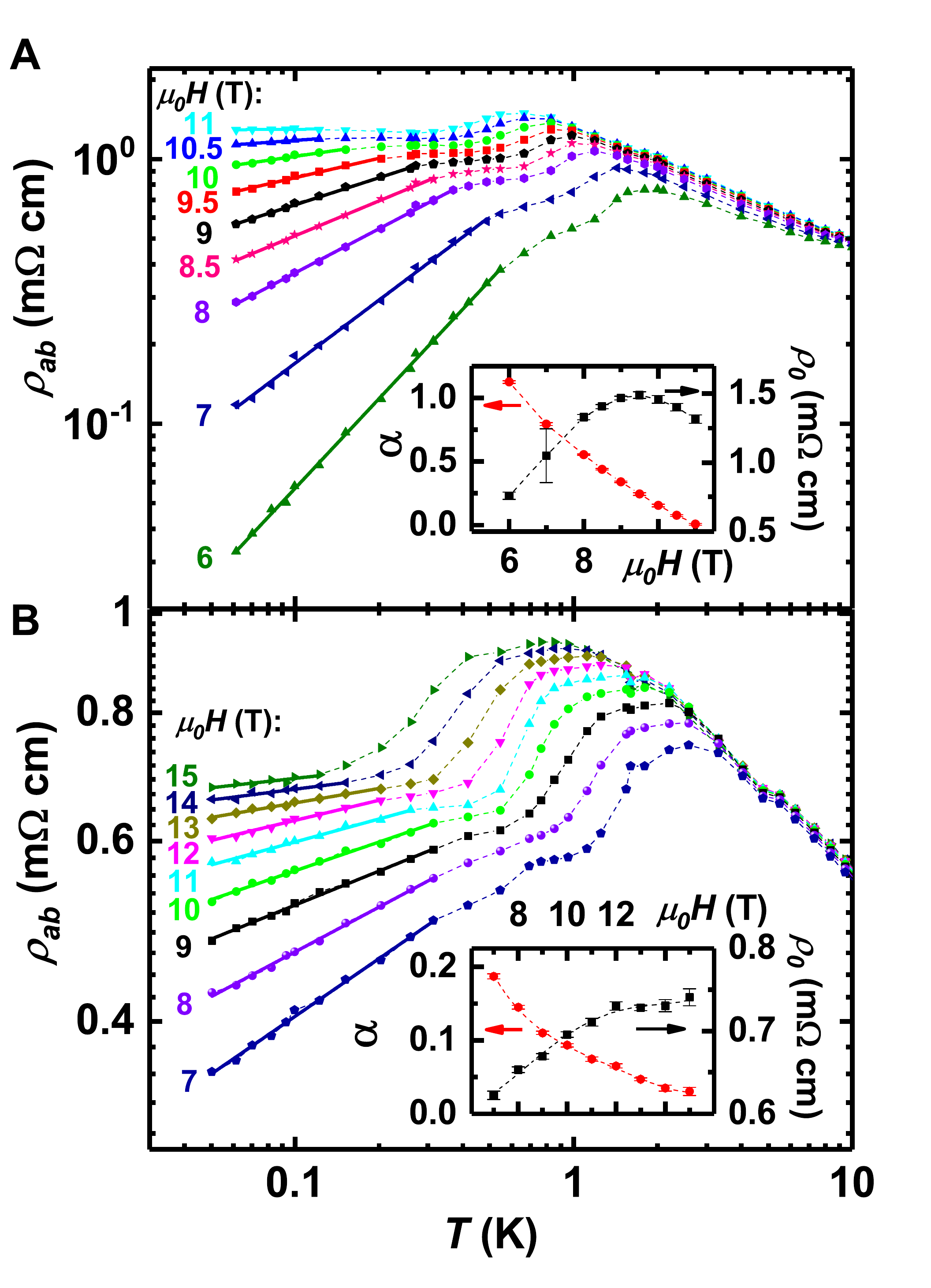}}
\caption{\textbf{The dependence of the in-plane resistivity on $\bm{T}$ at intermediate fields.} 
\textbf{A}, \lesco; \textbf{B}, \lnsco.  In both panels, ${\rho}_{ab}(T)$ for a given $H$, as shown, are plotted on a log-log scale.  Short-dashed lines guide the eye.  Solid lines represent power-law fits ${\rho}_{ab}(H,T) = {\rho}_{0}(H)T^{{\alpha}(H)}$.  Insets: Fitting parameters $\alpha(H)$ and ${\rho}_{0}(H)$.  Short-dashed lines guide the eye.  The linear resistance $R_{ab}\equiv\lim_{I_{dc}\rightarrow 0} V/I$ was measured with $I_{ac}\approx 10~\mu$A.   
}
\label{powerlaw}
\end{figure}

\clearpage

\begin{figure}
\centerline{\includegraphics[width=0.70\textwidth]{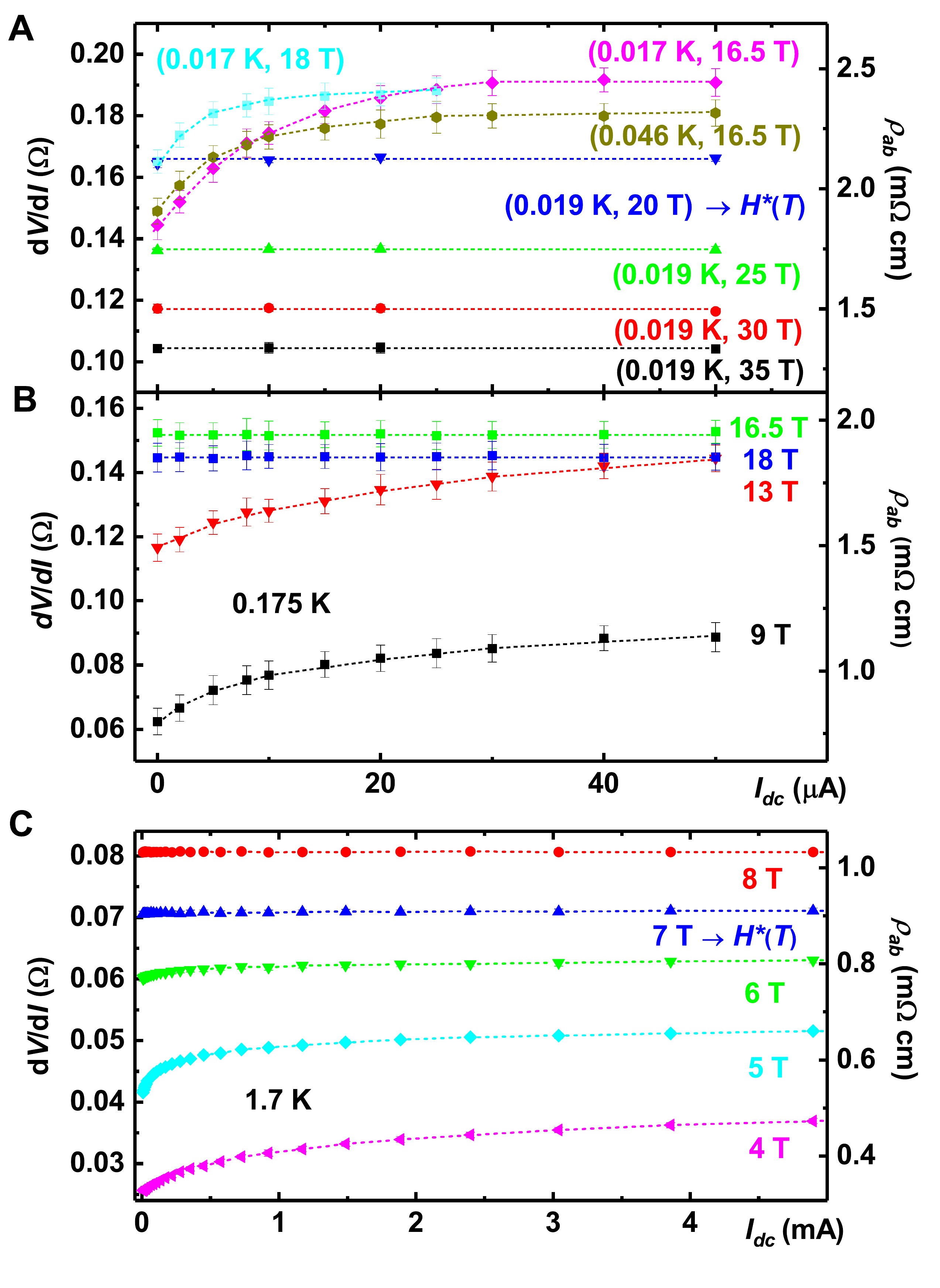}}
\caption{\textbf{Nonlinear in-plane transport in La$\bm{_{1.7}}$Eu$\bm{_{0.2}}$Sr$\bm{_{0.1}}$CuO$\bm{_4}$.} 
Representative traces of differential resistance $dV/dI$ vs. $I_{dc}$ for several $T$ and $H$, as shown. Changes of $dV/dI$ vs. $I_{dc}$ from non-Ohmic to Ohmic with increasing $H$ are shown at \textbf{A}, the lowest temperatures; \textbf{B}, 0.175 K, one decade higher than base $T$; and \textbf{C}, 1.7 K, two decades higher than base $T$. In \textbf{A} and \textbf{B}, the data were taken with $I_{ac}\approx 1~\mu$A, except for $20\leq H($T$)\leq 35$ in \textbf{A}, where $I_{ac}\approx 10~\mu$A. In \textbf{C}, the data were taken with $I_{ac}\approx 10~\mu$A, with $I_{dc}$ up to $\approx 5$~mA, where Joule heating was still absent. $H^{\ast}(T)$, determined as the magnetic fields where $dV/dI$ vs. $I_{dc}$ becomes Ohmic, are labeled in \textbf{A}, \textbf{B}, and \textbf{C}.  The uncertainties in $H^{\ast}(T)$, given in Fig.~2, reflect experimental resolutions in determining $H^{\ast}(T)$. Short-dashed lines guide the eye.
}
\label{FigS4}
\end{figure}

\clearpage

\begin{figure}
\centerline{\includegraphics[width=0.90\textwidth]{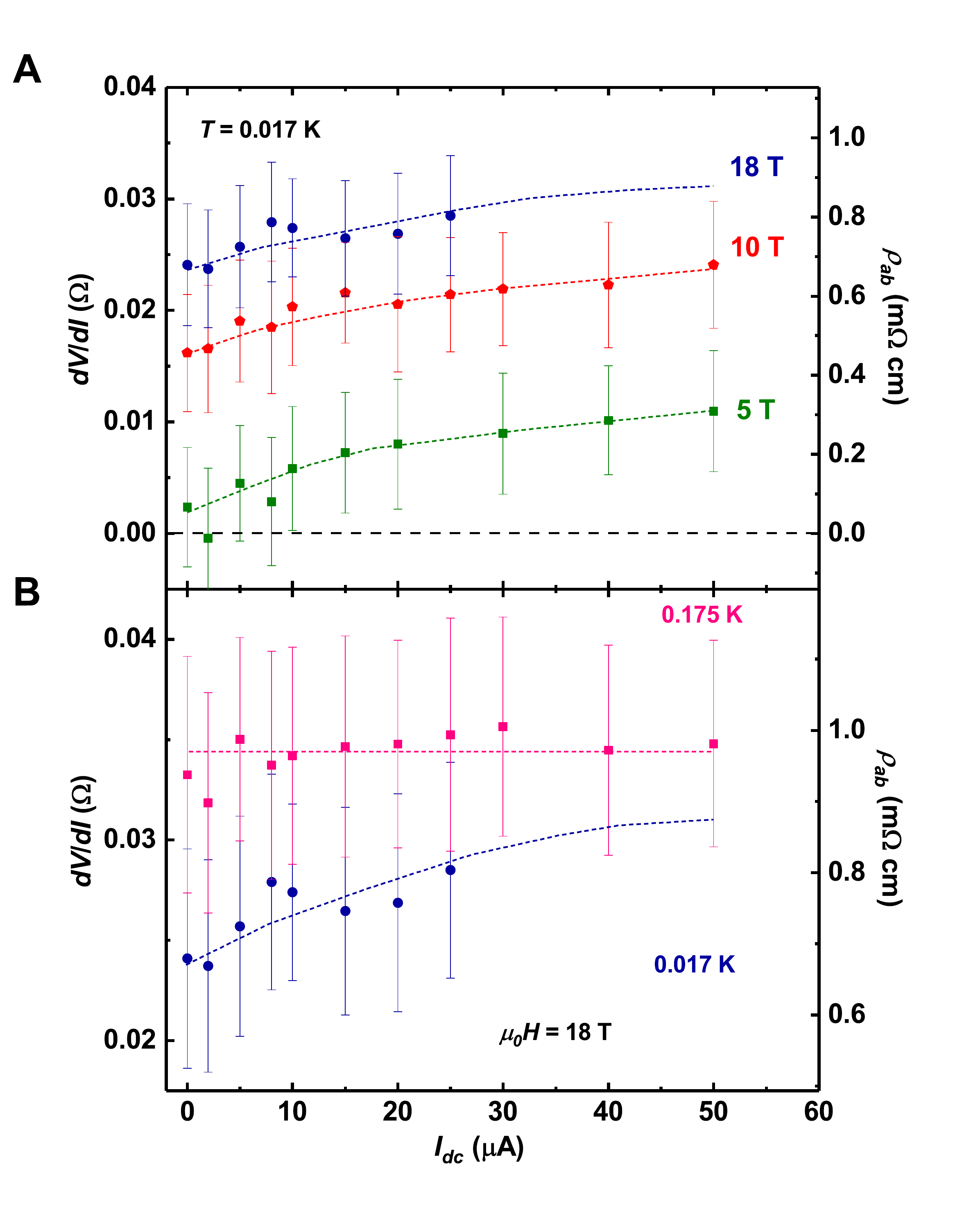}}
\caption{\textbf{Nonlinear in-plane transport in La$\bm{_{1.48}}$Nd$\bm{_{0.4}}$Sr$\bm{_{0.12}}$CuO$\bm{_4}$.} 
\textbf{A}, Differential resistance $dV/dI$ as a function of dc current $I_{dc}$ for several $H$, as shown, at $T=0.017$~K.  \textbf{B}, $dV/dI$ vs. $I_{dc}$ for $H=18$~T at two temperatures, as shown.  Short-dashed lines in (\textbf{A}) and (\textbf{B}) guide the eye.
}
\label{FigS5}
\end{figure}

\clearpage

\begin{figure}
\includegraphics[width=\textwidth]{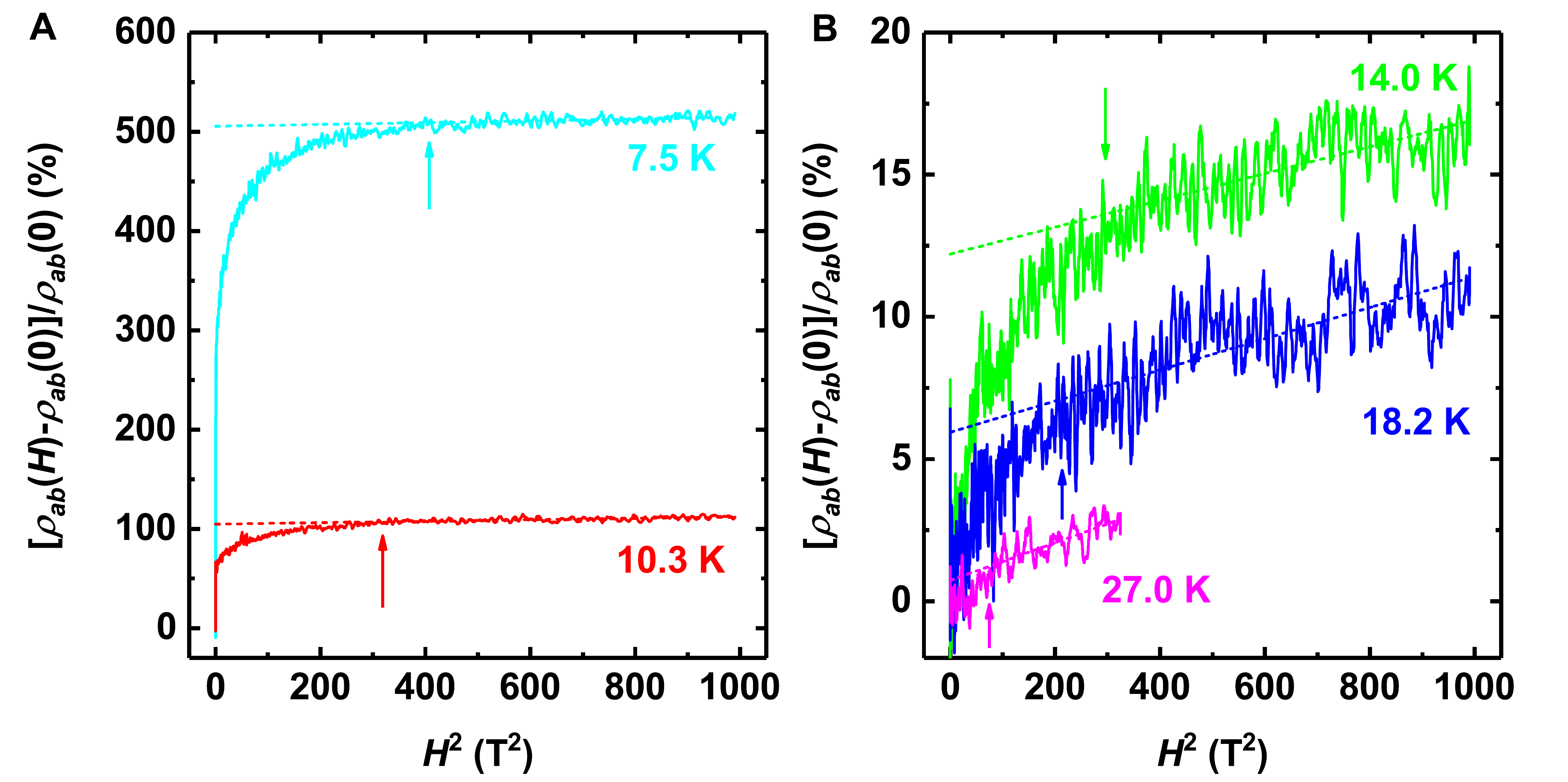}
\caption{\textbf{In-plane magnetoresistance of La$\bm{_{1.7}}$Eu$\bm{_{0.2}}$Sr$\bm{_{0.1}}$CuO$\bm{_4}$ vs. $\bm{H^{2}}$ for several $\bm{T>T_c(H=0)}$; $\bm{H\parallel c}$.}  In \textbf{A} and \textbf{B}, short-dashed lines are fits representing the contributions from normal state transport, i.e. they correspond to $[\rho_{ab}(H)-\rho_{ab}(0)]/\rho_{ab}(0)=\{[\rho_{ab}(0)]_{n}-\rho_{ab}(0)\}/\rho_{ab}(0)+{\alpha}H^{2}$.  The intercept of the short-dashed line shows the relative difference between the fitted normal state resistance and the measured resistance at zero field.  The difference between the short-dashed lines and the measured magnetoresistance is due to the superconducting contribution.  Arrows indicate $H_{c}'$, the field below which superconducting fluctuations become observable; it is determined as the field at which the deviation from the $H^2$ fit exceeds the noise level.  This method cannot be used at lower $T$ because the magnetoresistance develops a peak.}
\label{FigS6}
\end{figure}

\clearpage

%
\begin{figure}[t]
\centerline{\includegraphics[width=0.8\textwidth]{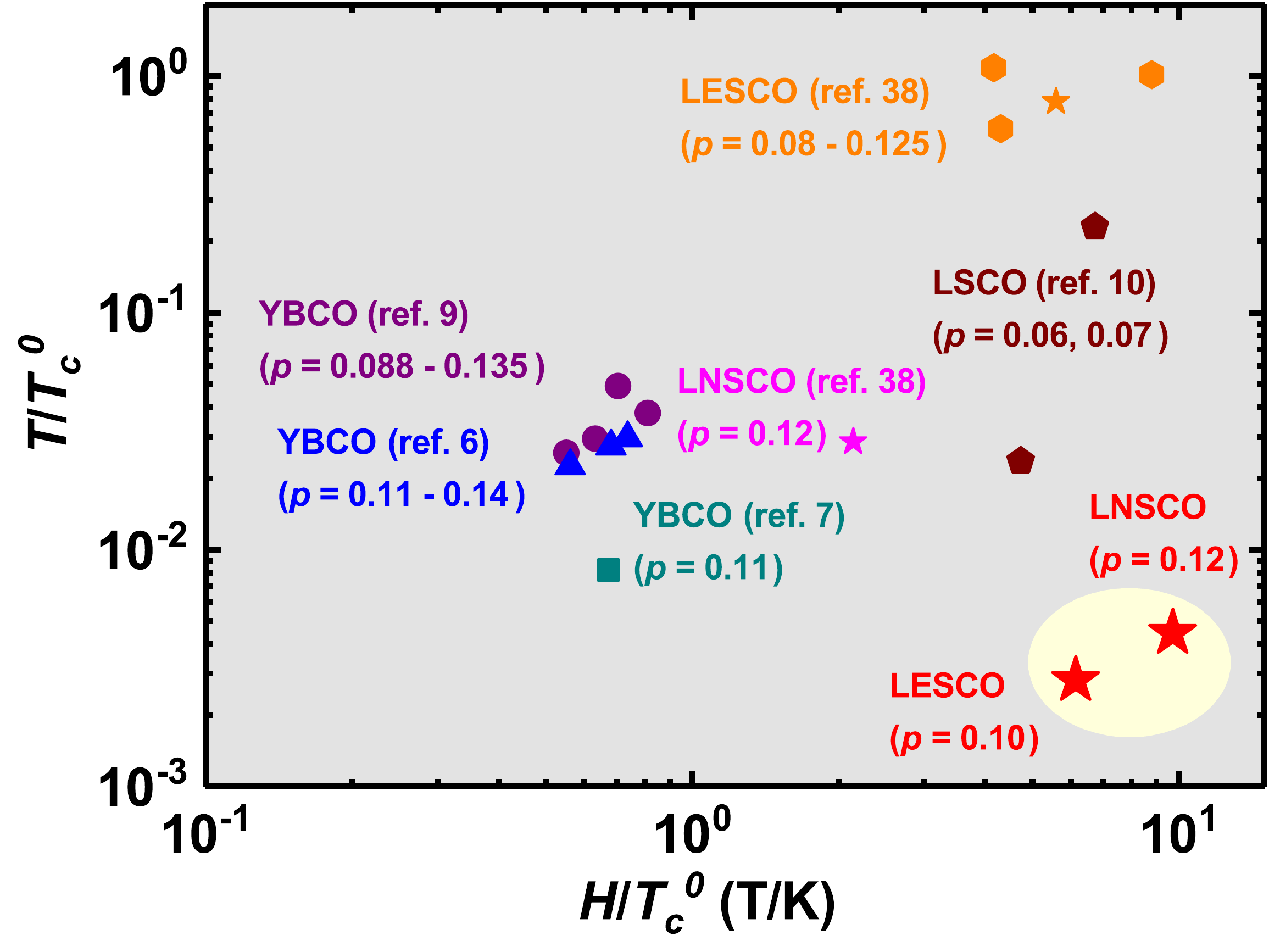}}
\caption{\textbf{Comparison of studies of upper critical field in various underdoped cuprates at different hole concentrations ($\bm{p}$).}  The symbols correspond to the lowest $T$ and highest $H$ reached in previous studies, as shown, normalized by $T_{c}^{0}$.  Our study (red star symbols for LESCO and LNSCO in the oval at lower right) clearly extends to effectively much lower $T$ and higher $H$ compared to earlier work on these materials at the same $p$ (star symbols) and other hole-doped cuprates.}
\label{sketch}
\end{figure}
%

\clearpage

%
\begin{figure}
\centerline{\includegraphics[width=0.78\textwidth]{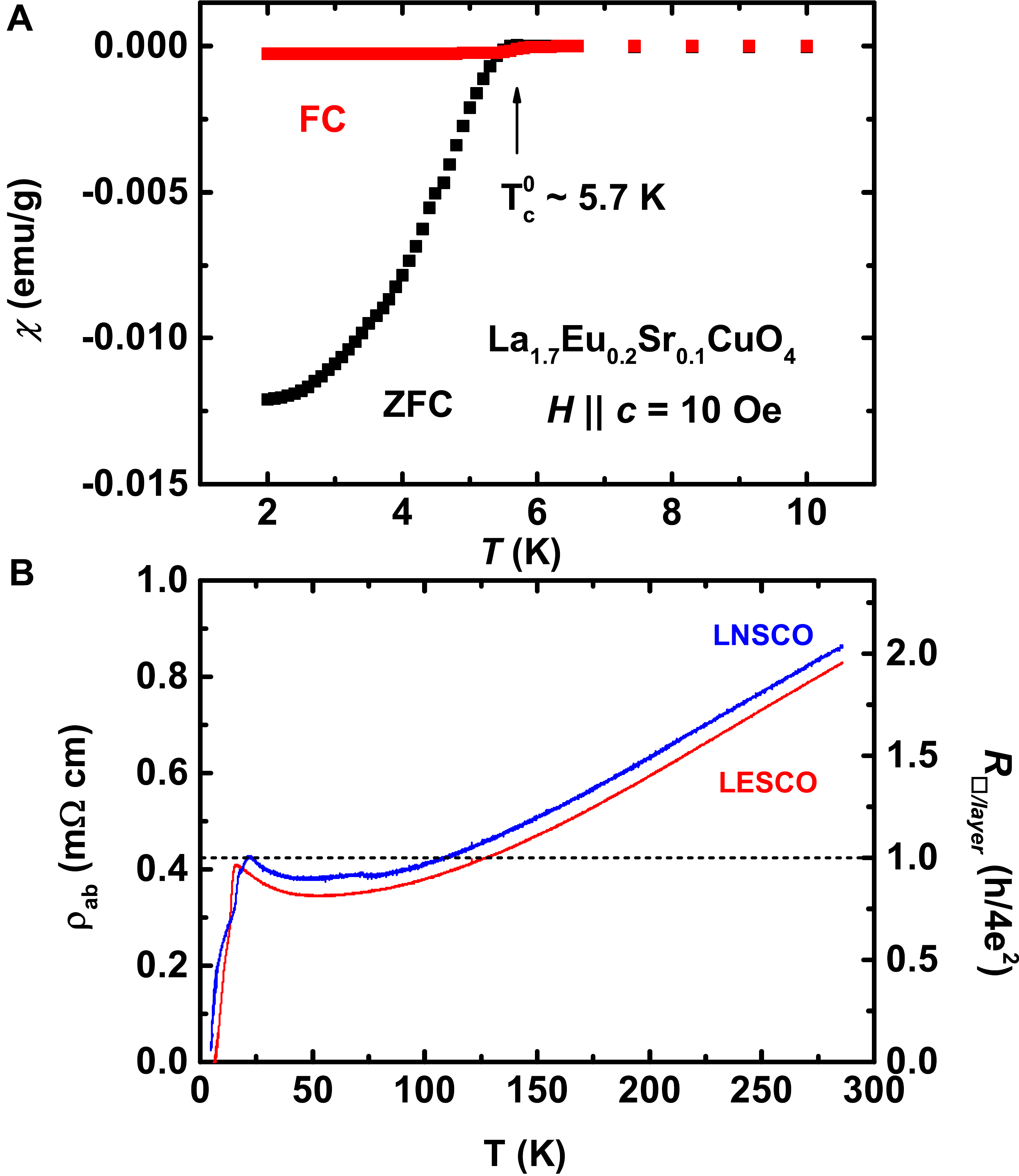}}
\caption{\textbf{Temperature dependence of the magnetic susceptibility and in-plane resistivity.} (\textbf{A}) The zero-field-cooled (ZFC) and field-cooled (FC) bulk magnetic susceptibility of \lesco. The SQUID measurement was performed on a \lesco\, crystal that was from the same batch, but different from the one on which the transport measurements were done. Nevertheless, $T_{c}^{0}\approx 5.7$ K, as identified from the onset of the bulk diamagnetic behavior, agrees very well with that obtained from our transport measurements, confirming the high quality of our crystals.  (See also Fig.~S10.) (\textbf{B}) The in-plane resistivity $\rho_{ab}$ of \lnsco  and \lesco; $H=0$.  The linear $\rho_{ab}(T)$ observed at high temperatures is higher in LNSCO single crystal sample than in LESCO, and it extrapolates to a higher residual ($T=0$) resistivity.  Therefore, both the residual resistivity and the normal-state resistivity, i.e. $\rho_{ab}$ at temperatures above the onset of superconducting fluctuations, may be used as a measure of disorder.  
}
\label{SQUID}
\end{figure}

\clearpage

\begin{figure}
\centerline{\includegraphics[width=0.8\textwidth]{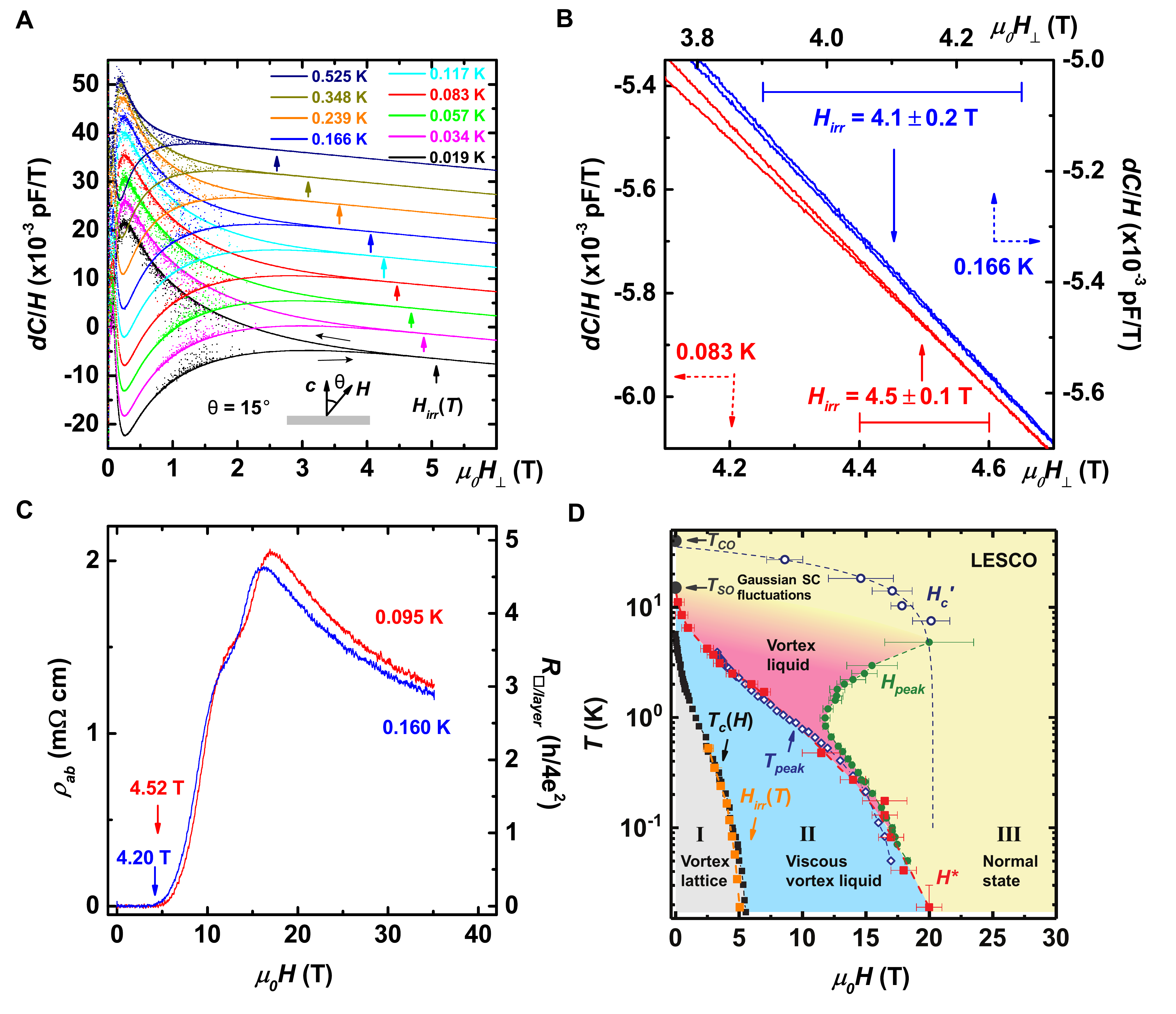}}
\caption{\textbf{Comparison of the irreversibility fields $\bm{H_{irr}(T)}$ with the resistive $\bm{T_{c}(H)}$ in \lesco.}  \textbf{A}, $dC/H$ vs. $\mu_{0}H_{\perp}$ for several $T$, as shown.  Except for the 0.019~K data, a constant shift of 0.005~pF/T is added from trace to trace for clarity.  The angle between $H$ and the crystalline $c$ axis (sketch) was $\theta$ = 15$^\circ$; $H_{\perp}=H\cos\theta$.  The capacitance $C$ was measured using an Andeen-Hagerling 2500A capacitance bridge, with a single crystal glued to a 1-mil-thick BeCu capacitive cantilever.  The same sample, different from the one used for transport measurements, was measured in Fig.~S9.  The capacitance change, $dC = C(H)-C(H=0)$, is proportional to the magnetic torque; $dC/H$ is proportional to the anisotropic magnetization [S8]. Arrows point at the onsets of bifurcation, i.e. $H_{irr}(T)$, determined as the fields where the hysteresis in dC exceeds the noise level $\sim 2 \times 10^{-5}$ pF. The linear change of $dC/H$ with $\mu_{0}H_{\perp}$ above $H_{irr}(T)$ reflects the paramagnetic Van Vleck contribution
[S9].  The noise below 2~T comes from superconducting magnets and does not affect the determination of $H_{irr}(T)$.  \textbf{B}, $dC/H$ vs. $\mu_{0}H_{\perp}$ near $H_{irr}(T)$ for 0.083 K (left, bottom axes) and 0.166 K (right, top axes). Solid arrows point at $H_{irr}(T)$. Horizontal bars represent the errors in $H_{irr}(T)$, reflecting the uncertainty in estimating the onsets of bifurcation. \textbf{C}, $\rho_{ab}(H)$ at $T=0.095$~K and 0.160~K close to those in \textbf{B}.  $T_{c}(H)$ are indicated by the arrows and associated values. \textbf{D}, $H_{irr}(T)$ added to the in-plane transport $T-H$ phase diagram with $H \parallel c$ axis.  The agreement between $H_{irr}$ and $T_c(H)$ obtained on two different samples confirms the high quality of our crystals.
}
\label{torque}
\end{figure}

\clearpage

\noindent\textbf{\large{{Supplementary Materials References}}}

\noindent [S1] M. Autore, P. Di Pietro, P. Calvani, U. Schade, S. Pyon, T. Takayama, H. Takagi, S. Lupi, Phase diagram and optical conductivity of La$_{1.8-x}$Eu$_{0.2}$Sr$_x$CuO$_4$.  \textit{Phys. Rev. B} {\bf 90}, 035102 (2014).\\
\noindent [S2] H.-H. Klauss, W. Wagener, M. Hillberg, W. Kopmann, H. Walf, F. J. Litterst, M. H\"ucker, B. B\"uchner, From antiferromagnetic order to static magnetic stripes: The phase diagram of $(La, Eu)_{2-x}$Sr$_x$CuO$_4$.  \textit{Phys. Rev. Lett.} {\bf 85}, 4590--4593 (2000).\\
\noindent [S3] J. Fink, V. Soltwisch, J. Geck, E. Schierle, E. Weschke, B. B\"uchner, Phase diagram of charge order in La$_{1.8-x}$Eu$_{0.2}$Sr$_x$CuO$_4$ from resonant soft x-ray diffraction.  \textit{Phys. Rev. B} {\bf 83}, 092503 (2011).\\
\noindent [S4]  M. K. Crawford, R. L. Harlow, E. M. McCarron, W. E. Farneth, J. D. Axe, H. Chou, Q. Huang, Lattice instabilities and the effect of copper-oxygen-sheet distortions on superconductivity in doped La$_2$CuO$_4$. \textit{Phys. Rev. B} \textbf{44}, 7749 (1991).\\
\noindent [S5]  N. Ichikawa, S. Uchida, J. M. Tranquada, T. Niem\"oller, P. M. Gehring, S.-H. Lee, J. R. Schneider, Local magnetic order vs superconductivity in a layered cuprate. \textit{Phys. Rev. Lett.} \textbf{85}, 1738 (2000).\\
\noindent [S6]   R. Daou, N. Doiron-Leyraud, D. LeBoeuf, S. Y. Li, F. Lalibert\'e, O. Cyr-Choini\`ere, Y. J. Jo, L. Balicas, J.-Q. Yan, J.-S. Zhou, J. B. Goodenough, L. Taillefer, Linear temperature dependence of resistivity and change in the Fermi surface at the pseudogap critical point of a high-$T_c$ superconductor.  \textit{Nat. Phys.} \textbf{5}, 31 (2009).\\
\noindent [S7] O. Cyr-Choini\`ere, R. Daou, J. Chang, F. Lalibert\'e, N. Doiron-Leyraud, D. LeBoeuf, Y. J. Jo, L. Balicas, J.-Q. Yan, J.-G. Cheng, J. S. Zhou, J. B. Goodenough, L. Taillefer, Zooming on the quantum critical point in Nd-LSCO.  \textit{Physica C} \textbf{470}, S12-S13 (2010).\\
\noindent [S8] L. Li,  Torque magnetometry in unconventional superconductors. Ph.D. thesis, Princeton University (2012).\\
\noindent [S9] M. H\"ucker, V. Kataev, J. Pommer, U. Ammerahl, A. Revcolevschi, J. M. Tranquada, B. B\"uchner, 
Dzyaloshinsky-Moriya spin canting in the low-temperature tetragonal phase of La$_{2-x-y}$Eu$_y$Sr$_x$CuO$_4$.  \textit{Phys. Rev. B} \textbf{70}, 214515 (2004).

\end{document}